\documentclass[journal = jcisd8, manuscript = article]{achemso}
\setkeys{acs}{usetitle = true, doi = true}
\usepackage{amsmath}
\usepackage{newtxtext,newtxmath}
\usepackage{xcolor}
\usepackage{booktabs}
\usepackage[version=4]{mhchem}
\usepackage{multirow}
\usepackage{xr-hyper}
\usepackage{hyperref}
\usepackage{bm}

\makeatletter
\newcommand*{\addFileDependency}[1]{
  \typeout{(#1)}
  \@addtofilelist{#1}
  \IfFileExists{#1}{}{\typeout{No file #1.}}
}
\makeatother

\newcommand*{\myexternaldocument}[1]{%
    \externaldocument{#1}%
    \addFileDependency{#1.tex}%
    \addFileDependency{#1.aux}%
}
\myexternaldocument{SI}

\newcommand{\angstrom}{\mbox{\normalfont\AA}}

\title{Separation of Flexible Enantiomers Using Shear Flow}
\author{Minh Nhat Pham}
\author{Levi Cherek}
\author{J. Daniel Gezelter}
\email{gezelter@nd.edu}
\affiliation[University of Notre Dame]{251 Nieuwland Science Hall, Department of Chemistry and Biochemistry, \\
University of Notre Dame, Notre Dame, Indiana 46556}

\begin{document}

\begin{tocentry}
\center\includegraphics[width=\linewidth]{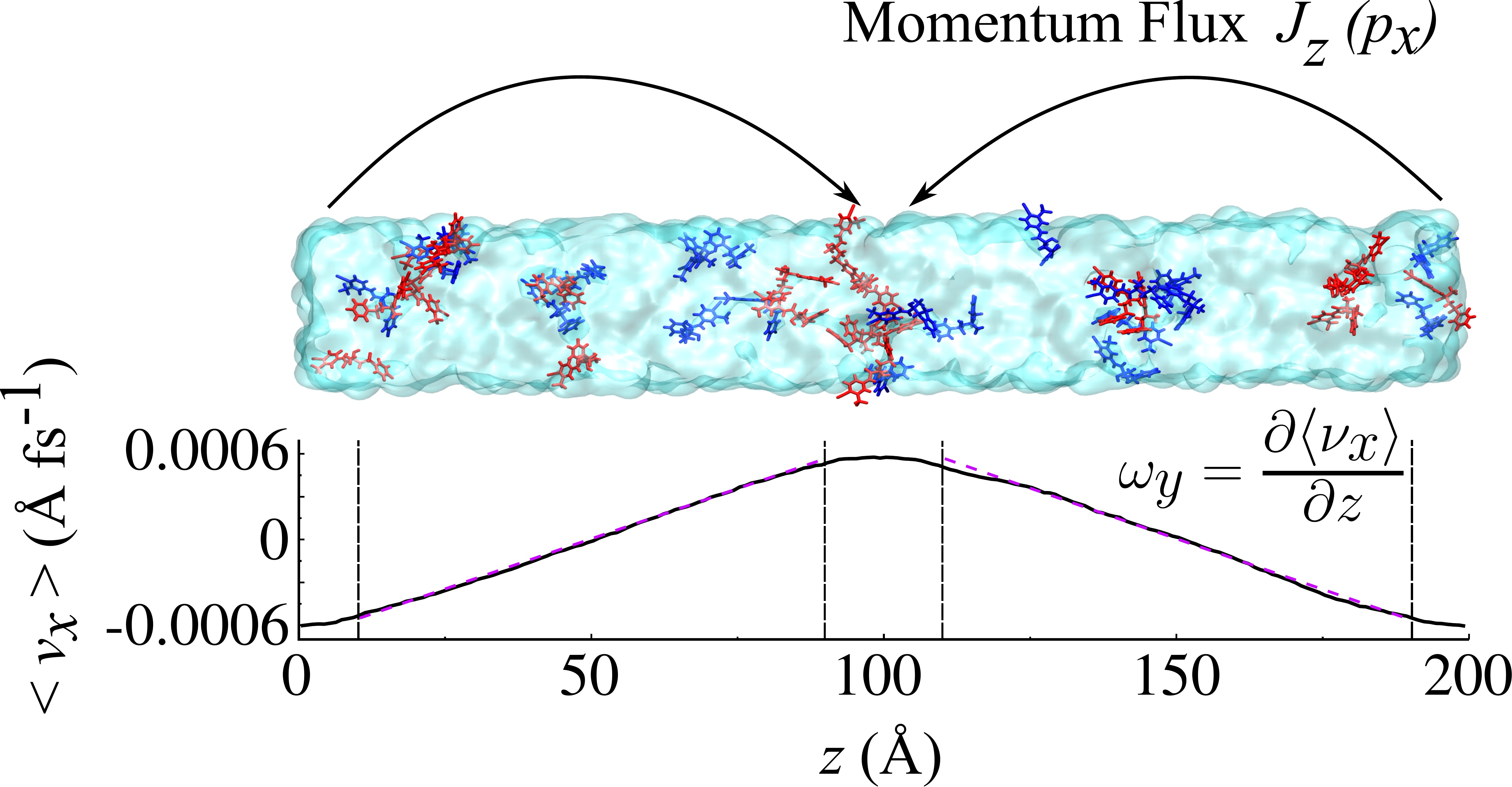} 
\end{tocentry}

\begin{abstract}
Mechanical separation of enantiomers is an attractive alternative to synthetic methods for producing enantiopure samples. Shear flow that produces solution vorticity has been shown to be a viable means for separating chiral objects on the micro- to nano-scale due to the tensorial nature of the interactions between chiral objects and the surrounding fluid. A recently-developed theory of molecular pitch characterizes these interactions using the resistance tensor and predicts the shear-induced separation of drug-like molecules from their optimized molecular geometries. We present a molecular dynamics study on the effects of incorporating molecular flexibility into the molecular pitch framework. We also evaluate the potential for enantiomeric separation of two drug molecules: bicalutamide (Casodex) and montelukast sodium (Singulair). Simulations reveal the emergence of flexibility-induced pitch distributions that result from conformational changes occurring in a realistic solvent environment. However, these distributions are  weakly 
influenced by the solvent identity and the shearing process, producing mean scalar pitch values that are close to those from optimized gas phase structures. Despite the opposing effect of translational diffusion at the molecular scale, racemic mixtures of flexible enantiomers show linear rates of separation at the 10 ns timescale, and we predict that cm-scale separation can be achieved within hours. Additionally, we provide estimates for parameters of a Taylor-Couette device for generating laminar shear flow, as well as considerations for future experiments.
\end{abstract}

\section{Introduction}

The separation of enantiomers has become increasingly important in pharmaceutical research.~\cite{Gorog1994,Rentsch2002,Carey2006,Nguyen2006,Agranat2012,Katzung2012,Todd2014} There are many chiral (left- or right-handed) molecules in which one enantiomer exhibits drug-like properties, while the other is either dangerous or ineffective. Examples of drugs with a dangerous enantiomer include ethambutol, an antitubercular agent whose  ($R,R$)-enantiomer causes optic neuropathy.\cite{Nguyen2006} In addition to addressing safety concerns of racemate drugs, the separation of pharmacologically active enantiomers from the inactive form has significant commercial implications. One notable example is the proton pump inhibitor omeprazole (marketed as Prilosec), which is a racemate, and esomeprazole (marketed as Nexium), the ($S$) enantiomer at the sulfur chiral center. The separation of the racemic mixture has allowed a `chiral switch' which permits re-development of a single enantiomer version of an already approved racemate drug. 

There has been significant progress in separating enantiomers after they have been formed by the primary synthetic reactions. Recently, a number of groups have discovered that relatively straightforward physical methods for separating enantiomers may be possible. These methods include the use of electric fields,\cite{Clemens2015} chiral resins,\cite{Shen2016} and most relevant to this work, vortex flows. Chiral environments that can be used to resolve enantiomers often involve chiral column chromatography,\cite{Gorog1994,Jacques1994,Nguyen2006} where enantiomers have distinct intermolecular interactions with a chiral stationary phase or with a chiral solvent.\cite{Nguyen2006,Katzung2012} The ability to reliably separate enantiomers using a physical process would make other expensive strategies, e.g. diastereomer-based  separations and tailored catalysts that produce enantiomeric excess,\cite{Jacques1994,Carey2006,Nguyen2006,Todd2014} less relevant to the pharmaceutical enterprise.

In a recent set of papers, Duraes and Gezelter developed a method for predicting the shear-flow separation of enantiomers using a set of hydrodynamic calculations on the structures of the molecules.\cite{Duraes2021,Duraes2023} Although the solvent itself may be achiral, shear-induced vorticity of the fluid introduces a dynamic chiral environment that can induce separation. Molecular rotation is induced by a velocity gradient in a fluid undergoing shear flow, and the coupling of the two enantiomers with the fluid then provides opposing propulsion through the fluid.

The potential separation of chiral objects by achiral fluid flow was first reported by Howard \emph{et al}.\cite{Howard1976,Hirschfelder1977} Inside a rotating drum, Howard and co-workers suspended dextro-tartaric acid crystals in Isopar H (isoparaffinic hydrocarbons) and observed pure macroscopic (chiral) crystals moving in specific directions depending on the vorticity of the fluid.\cite{Howard1976} The central idea of the Howard \textit{et al.} experiments has been extended to the separation of other small chiral objects. The scale of separable objects has been decreasing, from the millimeter scale,\cite{Makino2008,Hermans2015} to microfluidic devices,\cite{Marcos2009,Aristov2013,Ro2016} and most recently to nanometer-scale helicoids.\cite{Alcanzare2017} A theoretical model proposed by Tencer and Bielski suggested that the time required to resolve micrometer-scale or smaller chiral objects would be an insurmountable barrier,\cite{Tencer2011} but these experimental observations of  separation at small length scales has rekindled interest in shear-induced enantiomeric separation of molecules. Indeed, Hermans \emph{et al}.\cite{Hermans2019} emphasized the need for additional theoretical studies with intermolecular interactions treated with realistic physics-based modeling. 

Duraes and Gezelter developed a theory of pitch for chiral molecules interacting with a surrounding fluid.\cite{Duraes2023} They modeled rigid molecules as power screws (which are driven solely by an imposed torque supplied by the fluid).\cite{Bhandari2010,Shigley2016} If we consider a screw rotating with a (body-fixed) angular velocity vector ($\bm{\omega}_{b}$) in a medium, the resultant coupling with the medium can impart a velocity vector ($\mathbf{v}$). For an arbitrary rigid body, the relationship between $\bm{\omega}_{b}$ and $\mathbf{v}$ is mediated by a $3 \times 3$ \textit{pitch matrix}  denoted with the symbol $\mathsf{P}/2\pi$ in Eq. \eqref{eq:pitch_matrix_generalization} , 
\begin{equation}
\mathbf{v}=\frac{\mathsf{P}}{2\pi}\,\bm{\omega}_{b}
\label{eq:pitch_matrix_generalization}
\end{equation}
A rigid or semi-rigid molecule has both linear \textit{and} angular velocities, and frictional contributions to the motion of the body are described by a \textit{resistance tensor},\cite{Brenner1964,Garcia1980}
\begin{equation}
\renewcommand{\arraystretch}{0.7}
\left( \begin{array}{l}
 \mathbf{f} \\
 \mathbf{\tau}  \\
 \end{array} \right) =  - \left( \begin{array}{*{20}c}
   \Xi^{tt} & \Xi^{rt}  \\
   \Xi^{tr} & \Xi^{rr}  \\
\end{array} \right)\left( \begin{array}{l}
 \mathbf{v} \\
 \mathbf{\omega}_b \\
 \end{array} \right)~.
 \label{eq:resistance}
\end{equation}
In Eq. \eqref{eq:resistance}, $\Xi^{tt}$ and $\Xi^{rr}$ are $3 \times 3$ translational and rotational resistance (friction) tensors respectively, while $\Xi^{tr}$ is translation-rotation coupling tensor and $\Xi^{rt}$ is rotation-translation coupling tensor. When a particle moves in a fluid, it may experience a friction force ($\mathbf{f}$) and torque ($\mathbf{\tau}$) in opposition to the velocity ($\mathbf{v}$) and body-fixed angular velocity ($\mathbf{\omega}_b$).

 In Eqs. \eqref{eq:deriv1},\eqref{eq:deriv2}, and \eqref{eq:pitch_matrix_resistance_tensor}, we derive an analytical expression for the pitch matrix.  If we attribute the net force from translational motion entirely to the rotational contribution from the fluid,
\begin{equation}
    \Xi^{\mathrm{tt}}~\mathbf{v} = -\, \Xi^{\mathrm{rt}}~\bm{\omega}_{b}
    \label{eq:deriv1}
\end{equation}
we can then uncover the pitch matrix in terms of two blocks of the resistance tensor,
\begin{equation}
\Xi^{\mathrm{tt}}~\mathbf{v} = \Xi^{\mathrm{tt}} ~ \frac{\mathsf{P}}{2\pi}~ \bm{\omega}_{b} = -\, \Xi^{\mathrm{rt}}~ \bm{\omega}_{b}
\label{eq:deriv2}
\end{equation}
which implies 
\begin{equation}
\frac{\mathsf{P}}{2\pi} = - \, \left(\Xi^{\mathrm{tt}}\right)^{-1} \, \Xi^{\mathrm{rt}}
\label{eq:pitch_matrix_resistance_tensor}
\end{equation}
 $\mathsf{P}/2\pi$  is a $3 \times 3$ pitch matrix that mediates the relationship between the angular velocity $\bm{\omega}_{b}$ supplied by the fluid and the translational velocity $\mathbf{v}$ that is imparted to the molecule, \textit{i.e.}, Eq. \eqref{eq:pitch_matrix_generalization}. For a rigid body moving in a fluid flow with vorticity $\bm{\omega}\neq0$, the acquired angular velocity is known to be $\bm{\omega}_b=\bm{\omega}/2$.\cite{Duraes2021}

Rapid computation of molecular resistance tensors from atomic coordinates is now well established,\cite{Garcia1980,Duraes2021,Duraes2023} so it is possible to predict a molecule's `pitch' from its spatial structure, regardless of the viscosity of the surrounding liquid. The diagonal transformation of the pitch matrix at the \textit{center of pitch} (the special point at which the pitch matrix is symmetric) produces three \textit{moments of pitch} (eigenvalues) and three corresponding \textit{pitch axes} (eigenvectors). Each eigenvalue describes the resultant linear translation when the molecule rotates around the axis represented by the associated eigenvector. In Eq. (\ref{eq:pitch}), we define the \textit{scalar pitch} as a rotational invariant that takes into account contributions from all the pitch axes and provides an average estimation of the linear translation given a molecule's random orientations in a liquid,
\begin{equation}
    \frac{\left| P \right|}{2 \pi} = \sqrt{\frac{1}{3}\sum_i \lambda_i^2}
    \label{eq:pitch}
\end{equation}
where $\lambda_i$ is the $i^\mathrm{th}$ eigenvalue of the pitch matrix. Since the component blocks of the resistance tensors are linearly dependent on viscosity, the definition of the pitch matrix from Eq. \eqref{eq:pitch_matrix_resistance_tensor} suggests that molecular pitch is a structural property and is independent of viscosity. 

Given the importance of enantioseparation in the pharmaceutical industry, Duraes and Gezelter applied the pitch model to a library of 85 chiral drug molecules ~\cite{Kibar2014} and predicted their potential for shear-driven separation.\cite{Duraes2021} Their molecular pitch values were computed from optimized gas-phase geometries, and to evaluate the pitch model's potential to predict separation, they simulated racemic mixtures of rigid enantiomers under shear-driven vortex flow generated from application of velocity shearing and scaling (VSS) reverse nonequilibrium molecular dynamics (RNEMD).\cite{Kuang2012} Simulation results confirmed that vortex flow was capable of separating molecular enantiomers after a few nanoseconds, and numerical simulations of a coupled drift-diffusion equation predicted centimeter-length separation within hours.\cite{Duraes2021}

\subsection{Molecular Flexibility}
Despite some computational savings, rigid-body simulations ignore conformational changes and intramolecular interactions, which may be vital to a structural property like molecular pitch. Since different molecular geometries will produce different pitch matrices, it is imperative to study how molecular flexibility transforms this property. Additionally, the choice of solvent is relevant not only to the solubility of drug molecules, but also to the range of conformational changes exhibited by these molecules. We expect that solvation effects will have a significant impact on molecular pitch for flexible models of drug molecules. Molecules may also undergo conformational changes in response to the solvent shear. The potential for shear-induced conformational change is essential in determining whether shear flow can act as another pitch-modifying force. It is therefore  useful to assess  the overall practicality of shear flow separation of enantiomers incorporating molecular flexibility along with realistic simulation parameters, e.g. solvent identity and density.

Montelukast and bicalutamide were the two primary targets of this study, as their large predicted molecular pitches (0.325 and 0.246 \AA~/ rad respectively)\cite{Duraes2021} facilitate the observation of flexibility-dependent pitch modification. Montelukast (marketed as Singulair) is a leukotriene receptor antagonist. This chiral molecule is one of the most widely prescribed oral medications in the U.S. for the treatment and management of chronic asthma and recently allergic rhinitis.\cite{Nayak2007,Lee2020} Due to the low inherent water solubility of the free acid,\cite{Barbosa2016} montelukast is formulated and commercialized as a sodium salt to enhance its bio-availability. The traditional and current strategies to obtain enantiopure montelukast sodium include asymmetric synthesis with chemo-~\cite{Liang2010} or biocatalysts~\cite{Liang2010,Simic2022} and chromatography on amylose-based chiral columns.\cite{Liu1997,Vadagam2023} The asymmetric synthetic routes involve expensive keto-reducing agents, but montelukast sodium is still available as an enantiopure drug because the high prescription volume offsets the production costs. 

\begin{figure}[H]
    \centering
    \includegraphics[width=0.9\textwidth]{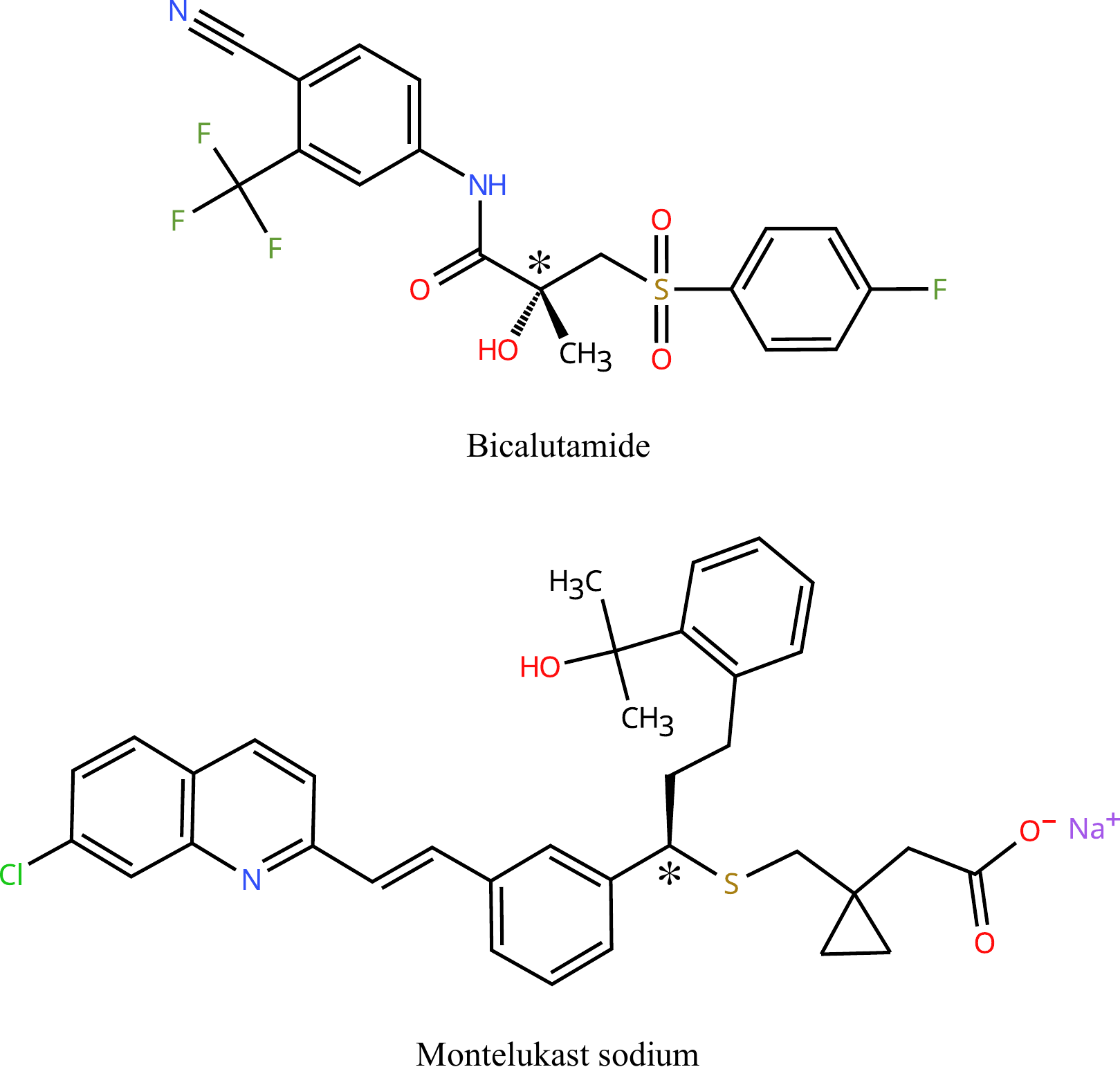}
    \caption{Structures of the active ($R$) enantiomers of bicalutamide (upper) and montelukast sodium (lower). The chiral centers are denoted with an asterisk (*).}
    \label{fig:drawn structures}
\end{figure}

Bicalutamide (marketed as Casodex) is a non-steroidal anti-androgen medication most commonly prescribed for localized prostate cancer.\cite{Masiello2002} Despite being marketed as a racemate, bicalutamide owes its therapeutic effects almost exclusively to the ($R$)  enantiomer,\cite{Tucker1988} while the non-toxic ($S$)-bicalutamide can be efficiently metabolized without adverse effects.\cite{Cockshott2004} Experimentally, enantiopure ($R$)-bicalutamide can be acquired through asymmetric synthesis~\cite{James2002} or post-synthetic chiral chromatography,\cite{Sadutto2016} but the current methods are cost-ineffective, chemically elaborate, or do not scale well (in the case of chromatography). Given the molecular structures in Fig. \ref{fig:drawn structures}, a significant number of rotatable bonds give both target molecules a degree of flexibility suitable for a practical study on molecular pitch. Additionally, their commercial values as therapeutics can aid in motivating future experimental attempts to utilize shear-driven vortex flow as a cost-effective method of enantiomer separation in pharmaceutical settings. 

\section{Methods}
Two types of molecular dynamics (MD) simulations were the main investigative methods of this study. The first set of simulations surveyed the conformational landscape of flexible bicalutamide and montelukast sodium (both as the inactive ($S$) enantiomers) in different solvents. These were carried out as micro-canonical (NVE) simulations of a single solute in explicit solvent boxes. The target molecule's geometry was sampled every 1 ps throughout a 10-ns simulation. Specifically, the chosen solvents were SPC/E water, ethanol, benzene, acetone (bicalutamide only), and methanol (montelukast sodium only). 

\begin{figure}[H]
    \centering
    \includegraphics[width=\textwidth]{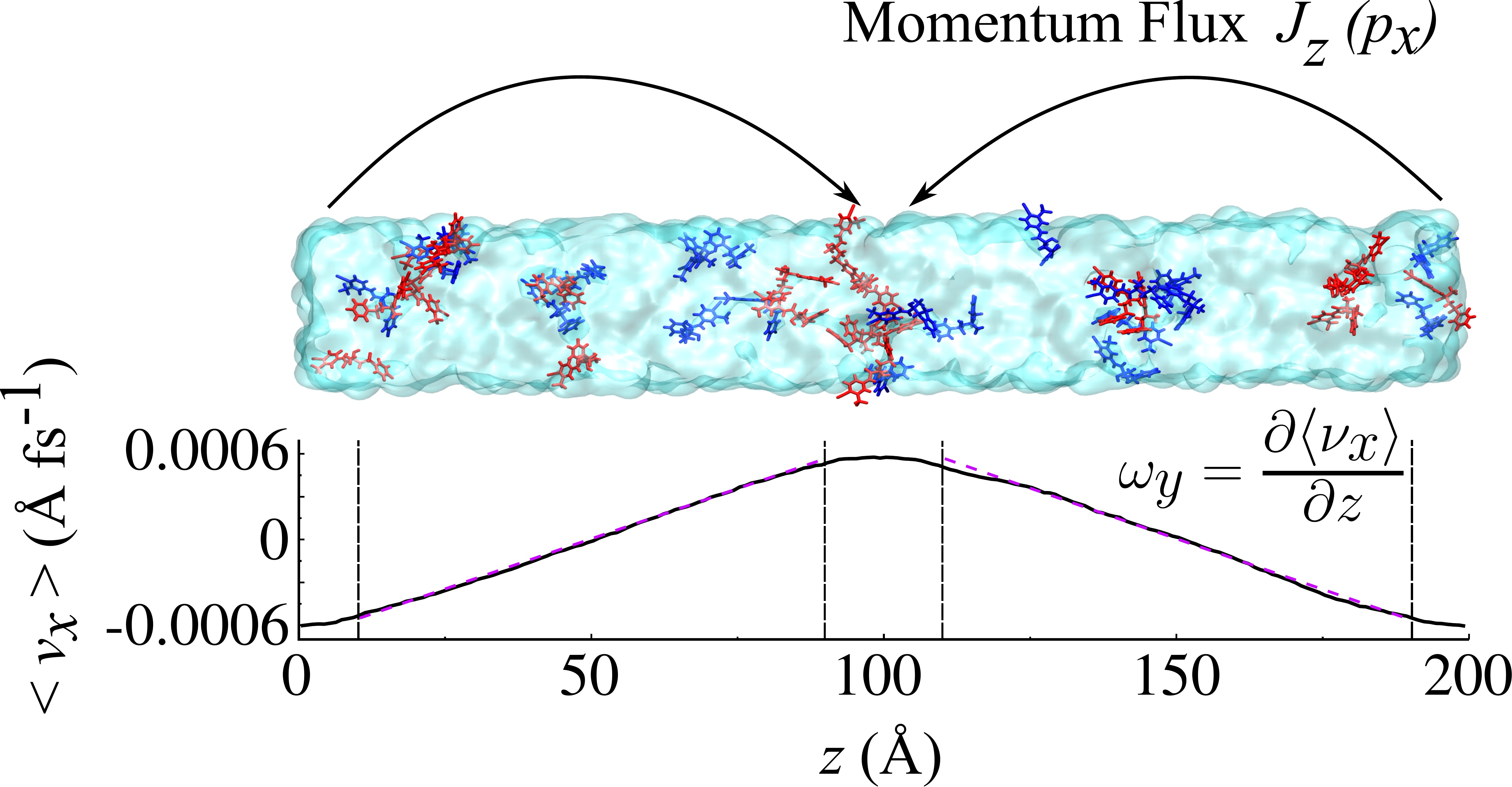}
    \caption{A schematic of a simulation cell undergoing shear through an imposed momentum flux. The flux was applied using OpenMD's velocity shearing and scaling reverse non-equilibrium molecular dynamics (VSS-RNEMD) algorithm. Above: A side view of the simulation box containing a racemic mixture of bicalutamide (shaded by red and blue colors) in acetone. The momentum flux $J_Z (p_x)$ was imposed on all atoms in the exchange regions (bounded by divided lines and connected by the arrows). Below: Resulting velocity gradients in the $x$ direction formed inside the in-between RNEMD regions. Their vorticities point in opposite directions. Linear fits of the vorticity are denoted by purple dotted lines.}
    \label{fig:simulationcell}
\end{figure}

A second set of simulations studied how racemic mixtures of the flexible solutes respond to an imposed momentum flux (Fig. \ref{fig:simulationcell}). These simulations are carried out using the velocity shearing and scaling (VSS) variant of reverse non-equilibrium molecular dynamics (RNEMD).\cite{Kuang2012}  The imposed momentum flux is an external shear stress, and the system responds by creating a momentum gradient in the simulation cell. After coming to a steady state, the simulation cell experiences two regions of vortex flow that rotate the solute molecules. Data collection included snapshots of molecular geometries taken every 10 ps and $y$-axis displacement correlation functions, $\langle \delta y(t) \rangle$ of the solute molecules.

Although the velocity profile shown in Fig. \ref{fig:simulationcell} resembles Poiseuille flow in a pipe or capillary, there are important differences. First, the VSS-RNEMD algorithm preserves total linear momentum in the system, so there is no net flow. Second, there are no walls, and no regions where the fluid is experiencing stick boundary conditions. Most importantly, in the regions between the two RNEMD exchange regions, the velocity profile is linear rather than quadratic. The RNEMD algorithm imposes a uniform shear stress, and under Newton's law of viscosity, the response is a uniform velocity gradient in a fluid.

In all cases, five (5) replicas of each simulation were generated using random resampling of molecular placements and random re-seeding of atomic velocities prior to the first equilibration stage. In the sections that follow, all figures and data presented in tables represent the mean of the five replicas, and 95\% confidence intervals\cite{Riley2006} around these means were calculated using $\varepsilon  = 1.96 \sigma / \sqrt{N-1}$, where $N$ is the number of replicas, and $\sigma$ is the standard deviation of the sample mean. 

\subsection{Force field}
The solutes and most solvent molecules were modeled using parameters from the General AMBER Force Field version 2 (GAFF2).\cite{Wang2004}  Partial charges were assigned using the AM1-BCC semi-empirical charge model.\cite{Jakalian2002,Silva2012,Kagami2023}  The partial charges on methanol and ethanol were adapted from the work of Fennell, Wymer, and Mobley,\cite{Fennell2014} which provides accurate descriptions of solvation free energies for alcohols. Rigid SPC/E water was used as the solvent in all aqueous systems.\cite{Berendsen1987}  Simulations were carried out using the real space damped shifted force (DSF) approach to compute electrostatic interactions with a cutoff of 12 \AA~ and a damping parameter $\alpha = 0.18~\angstrom^{-1}$.\cite{Fennell2006} 

In addition to the sodium salt, montelukast was also modeled as a free acid to aid in the assessment of electrostatic interference during separation.
Simulation protocols and all unique force field parameters (those not widely released as part of GAFF2) and partial charges used in this work are given in the Supporting Information (SI).

\section{Results and Discussion}

\subsection{Solvent-dependent distributions of molecular pitch}

\begin{figure}[H]
    \centering
    \includegraphics[width=\textwidth]{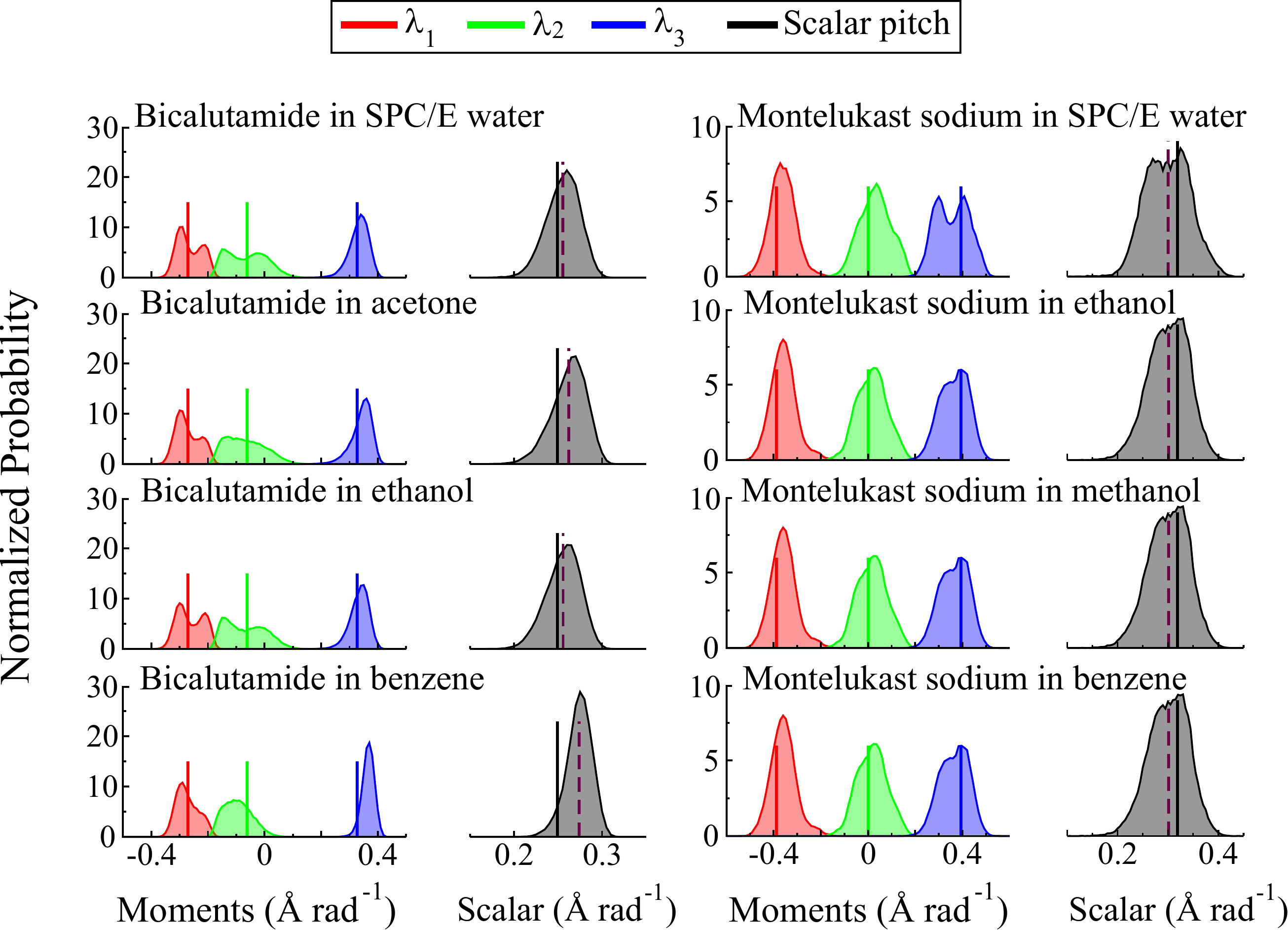}
    \caption{Distributions of the three moments of pitch $\lambda_1, \lambda_2, \lambda_3$ and the scalar pitch, $|P|/2\pi$, taken from flexible ($S$)-enantiomers of bicalutamide (left) or montelukast sodium (right) in various solvents. Pitch values for the optimized gas phase structure are indicated with vertical lines, and the mean scalar pitch for the flexible molecules is indicated with a dashed line.}
    \label{fig:casodex singulairNa 4solvents}
\end{figure}

Flexibility allows the molecules to adopt different conformations, each interacting with the surrounding solvent and creating a dynamic molecular pitch matrix. For each of the drug-solvent pairings that were simulated, we constructed normalized probability distributions of the scalar pitch as well as the three moments of pitch. In Fig. \ref{fig:casodex singulairNa 4solvents}, we show distributions of the molecular pitch captured from 50,000 instantaneous solution-phase geometries of the ($S$) enantiomers and compare these to the values of the optimized (gas-phase) geometries. Also shown is the mean value of the scalar pitch from the distribution of flexible molecules. 

\begin{figure}[H]
    \centering
    \includegraphics[width=\textwidth]{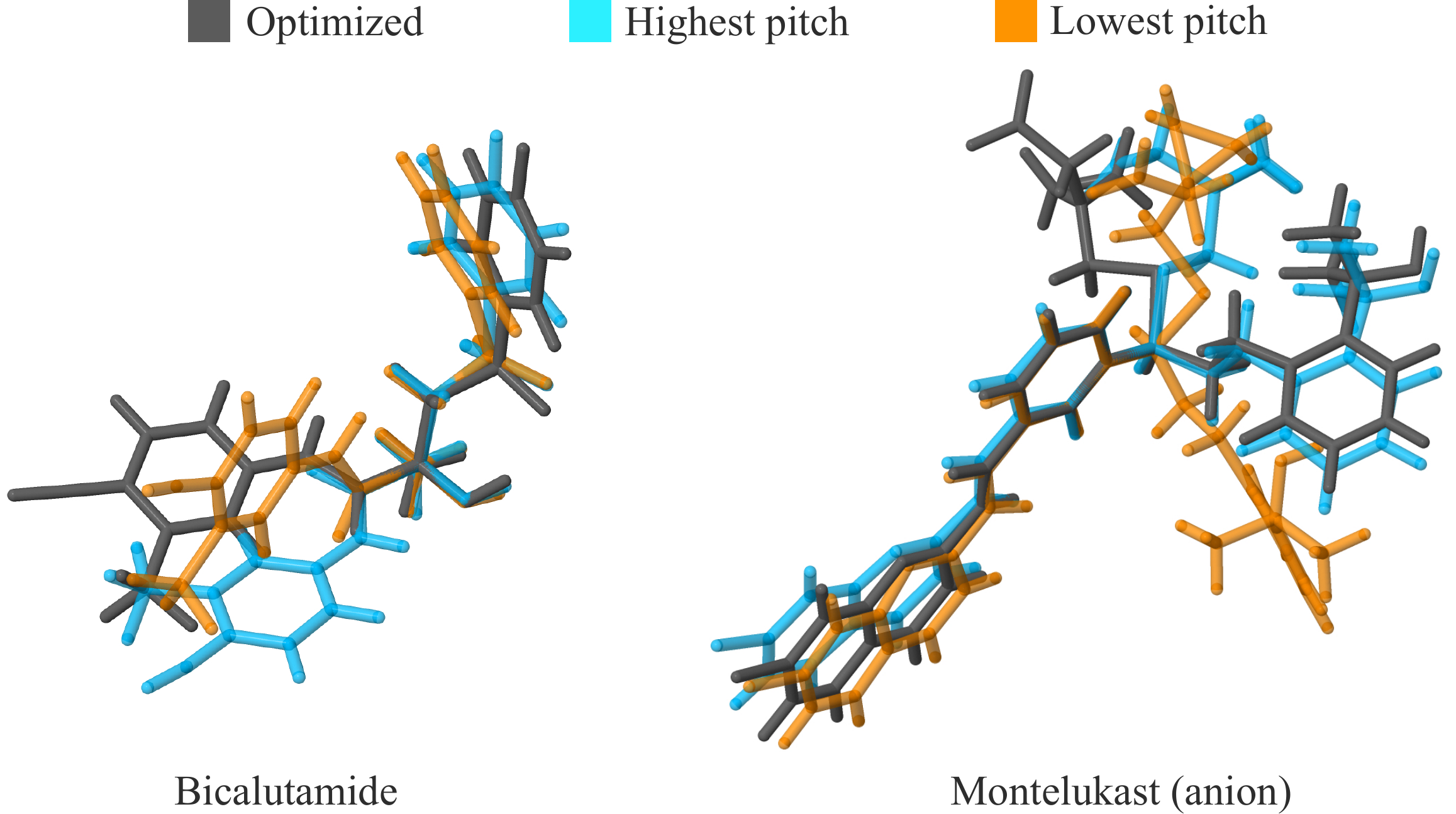}
    \caption{Molecular structures of ($S$)-bicalutamide (left) and the anion of ($S$)-montelukast sodium (right). The gas phase optimized geometries (dark gray) are superimposed with  the structures with the lowest (orange) and highest (blue) scalar pitch values. Solvent-mediated conformational changes lead to some of the observed variations in distributions of molecular pitch values.}
    \label{fig:structures}
\end{figure}

For bicalutamide (Casodex), the mean of the flexible distribution of molecules exhibits an increased scalar pitch relative to the optimized gas-phase geometry, while for montelukast sodium, flexibility reduces the molecular pitch. This observation holds for all of the solvents studied, and it does not appear to be related to molecular solubility in those solvents. Bicalutamide is generally not soluble in water,\cite{Cockshott2004} slightly soluble in ethanol,\cite{Le2009} and soluble in acetone.\cite{Cheng2023} It is a relatively small drug molecule with a significant fraction of polar terminal groups, so the low solubility likely arises from strong intermolecular interactions in the crystalline form. When solvated, bicalutamide exhibits a perturbed angle between two arms of the molecule but undergoes no other significant conformational changes. From the distributions of the moments of pitch, the variations between different solvents suggested the emergence of solvent-mediated conformations. In Fig. \ref{fig:structures}, it is clear that bicalutamide experiences a reduced scalar pitch when the two arms form a smaller angle relative to the optimized geometry. The skew towards higher scalar pitch values observed in benzene could thus be explained by a higher probability of the ``open'' conformations, where the two arms have a larger degree of separation. Bicalutamide crystallizes in two major conformations where the angle between the two arms is the major conformational change.\cite{Vega2007}  Our observations of the ``open'' conformations in benzene also agrees with previous findings that benzene is able to induce the transition from ``closed'' to ``open'' structures in bicalutamide through  stronger intermolecular interactions with the open structure.\cite{Mololina2025}  We also note that the ``closed'' conformation was not detected in any simulation snapshots with the other (polar) solvents. 

In contrast, montelukast sodium (the sodium salt of Singulair) is freely soluble in ethanol, methanol, and water.\cite{Barbosa2016,Oneil2013} The polar terminal groups are at the extremities of each of the three arms surrounding a largely hydrophobic core. In polar solvents, one would expect compression or folding of the structure to bury the hydrophobic core. 
The optimized gas phase geometries and structures that exhibit both high and low scalar pitches are shown in Fig. \ref{fig:structures}. Contrary to expectations, the majority of conformations exhibited by montelukast sodium had scalar pitch values concentrating around that of the optimized structure, which suggested the preservation of the high-pitch molecular frame close to the most stable gas phase conformation. For the anionic montelukast ion, molecular pitch was significantly reduced when the flatter three-armed propeller optimized conformation collapsed due to the high flexibility of the two carboxylate-containing and hydroxyl-containing arms. Consequently, different solvents, including SPC/E water, did not produce significant changes in the pitch distributions.

The higher variability in scalar pitch distributions of montelukast compared to those of bicalutamide underlines a central difference in the degree of flexibility between the two molecules. The larger and more flexible montelukast sodium structure allows it to adopt a wider range of conformational changes, which might lead to a broader pitch distribution and a reduced mean scalar pitch. In general, solvent interactions did not significantly alter the molecular pitch distributions in these molecules, while inherent molecular flexibility dictated most of the observed distributions around gas phase structures. Some conformations exhibit much higher or lower scalar pitches than the mean values, but those require very specific and short-lived spatial arrangements. Conformations with higher pitch values tend to differ marginally from the optimized structure, as seen in Fig. \ref{fig:structures}. 


\subsection{Flexible molecules in solutions undergoing shear}

\begin{figure}[H]
    \centering
    \includegraphics[width=\textwidth]{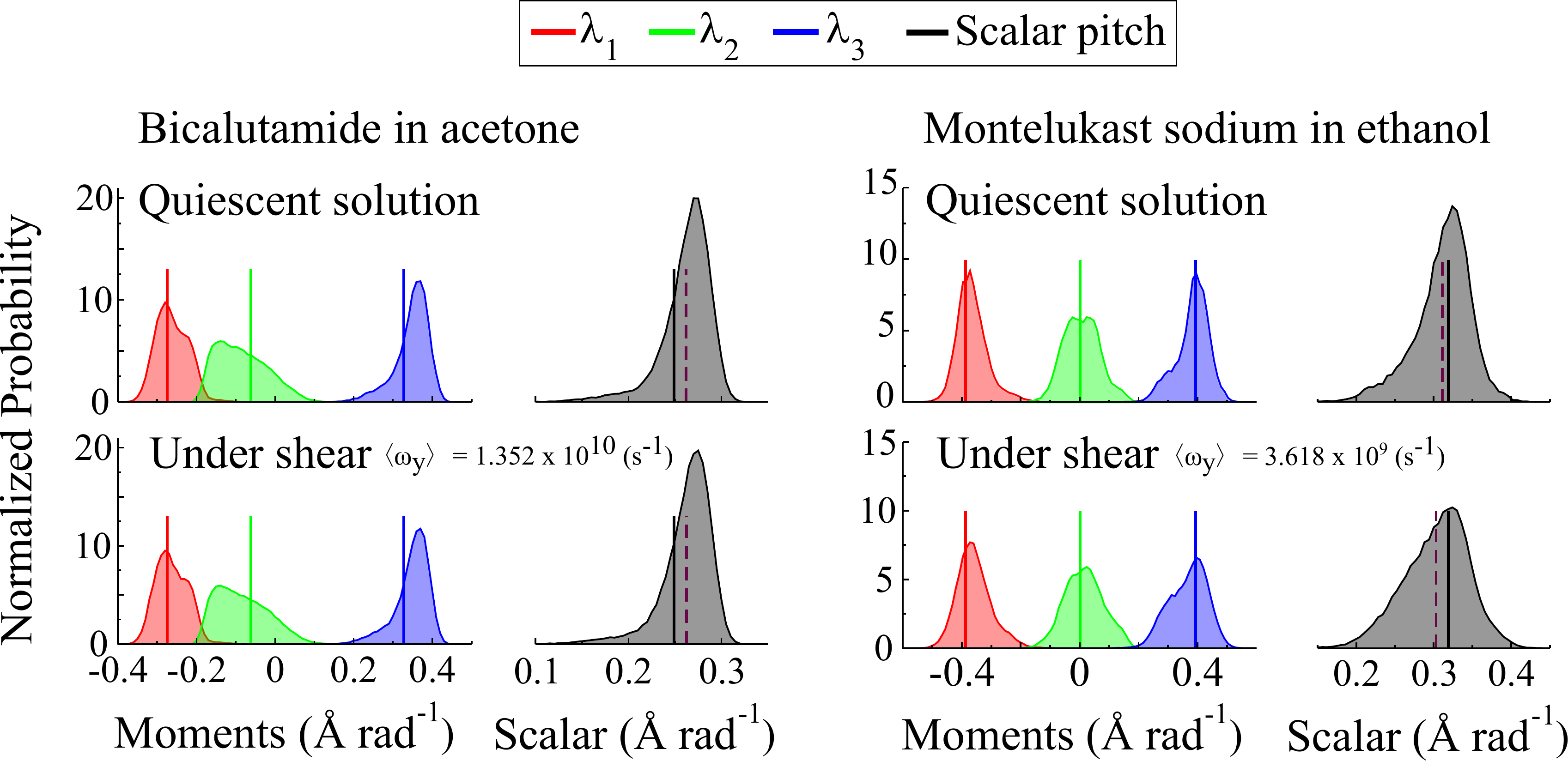}
    \caption{Distributions of the three moments of pitch $\lambda_1, \lambda_2, \lambda_3$ and the scalar pitch, $|P|/2\pi$, taken from ($S$)-enantiomers of bicalutamide (left) or the montelukast sodium (right) in their appropriate solvents. Both the quiescent and shearing solutions contain a racemic mixture of each target molecule. Pitch values for the optimized gas phase structure are indicated with vertical lines, and the mean scalar pitch for the flexible molecules is indicated with a dashed line.}
    \label{fig:casodex singulairNa shear vs no-shear}
\end{figure}

To better understand the effects of physical shearing on the molecular pitch of flexible molecules, we studied the pitch distributions of the ($S$) enantiomers of the two target molecules in solutions undergoing shear, as shown in Fig. \ref{fig:casodex singulairNa shear vs no-shear}. For this purpose, the enantiomers are simulated along with their mirror-image counterparts in a racemic mixture dissolved in appropriate solvent boxes. A velocity shearing and scaling (VSS) reverse nonequilibrium molecular dynamics (RNEMD) method~\cite{Kuang2012} was utilized to carry out shearing simulations. The simulation cell has one long axis $(z)$ and $x$-axis velocity exchanges are carried out between two RNEMD regions separated in $z$.  Over time, these velocity exchanges produce a velocity field with two regions of uniform vorticity $\omega_y = \partial \langle v_x \rangle /\partial z$.  Details on simulation box parameters and RNEMD regions for each molecule can be found in the SI.  

With a higher concentration of solute molecules, solubility and solute-solute interactions become more relevant to conformational landscapes. 
Being under shear does not alter the overall pitch distributions of ($S$)-bicalutamide. 
In contrast, ($S$)-montelukast sodium showed an increased probability of conformations whose scalar pitch values were below 0.3 \AA / rad during 5 ns of shearing. Shear-induced variations in the pitch distributions were likely caused by bringing the three molecular arms into closer contact during molecular rotation.
As a larger and more flexible molecule, montelukast sodium seems to be more strongly affected by the shearing process, and its mean scalar pitch was generally reduced. 
However, given the small impact of shearing on bicalutamide, we are unable to draw more general conclusions about the impact of shearing on molecular pitch.
We have performed a preliminary study using three different values of imposed solution vorticity, \textit{i.e.}, different shearing rates, to compare the molecular pitch distributions of ($S$)-enantiomers of free acid montelukast. This data can be found in the SI. In concert with the results in Fig. \ref{fig:casodex singulairNa shear vs no-shear}, shear flow appears to have negligible effects on altering pitch moments even if it acts to reduce scalar pitch in the case of montelukast sodium. 


\subsection{Practical separation of flexible molecules}

Using rigid-body bead models, Duraes and Gezelter showed that enantiomeric separation was possible under local shear flow for a wide range of molecules and solvents.\cite{Duraes2021} However, rigid-body simulations do not capture the conformational changes and subtle intermolecular interactions present in flexible models, which are crucial for assessing the practicality of the predictive molecule pitch model. Therefore, we extended these simulations to flexible bicalutamide and montelukast sodium molecules while keeping simulation conditions such as solution concentration and density close to realistic values. 

\begin{figure}[!ht]
    \centering
    \includegraphics[width=0.9\textwidth]{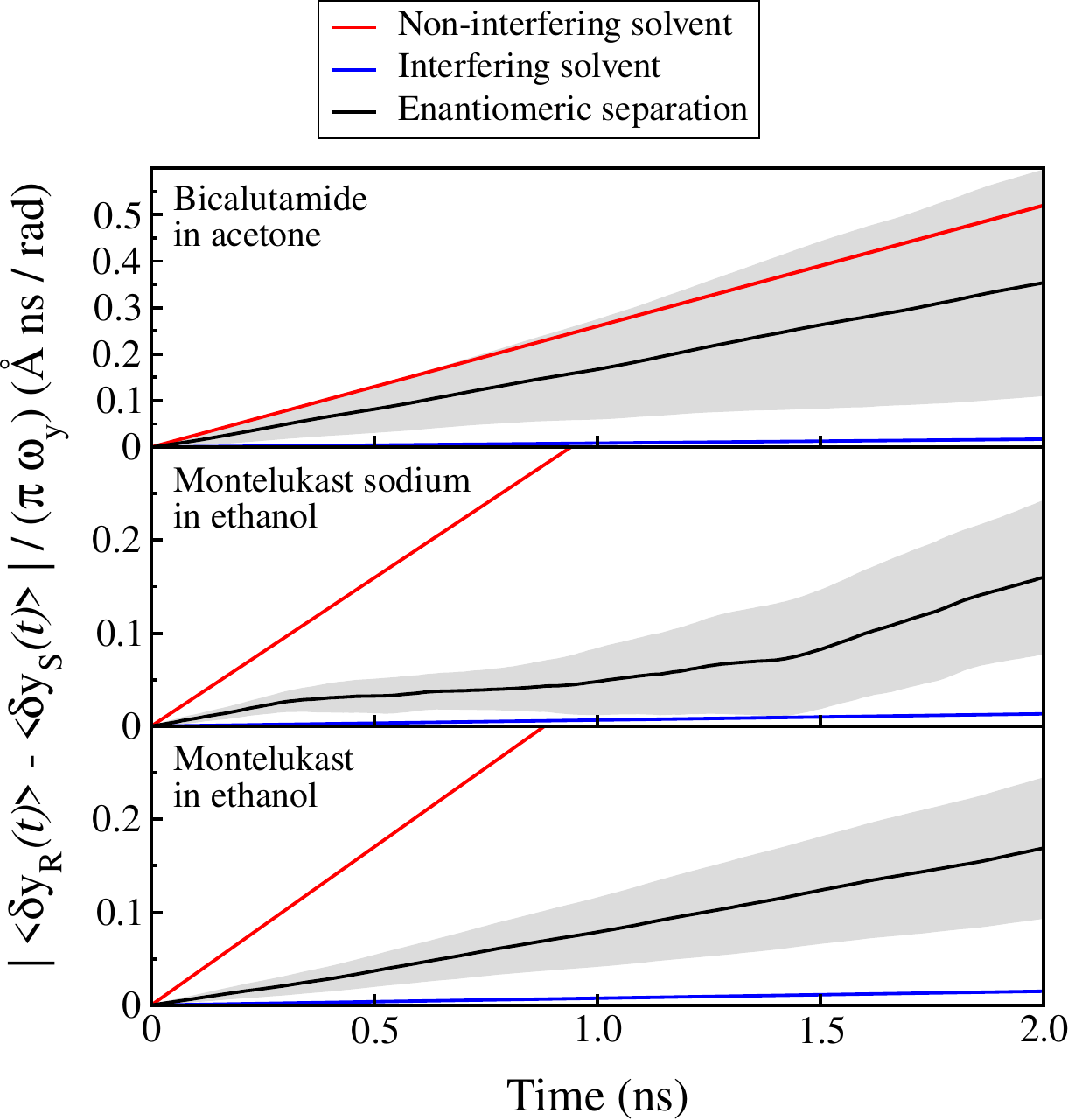}
    \caption{Mean enantiomeric separation (black lines) for a range of chiral molecules and solvents simulated under conditions of constant shear vorticity ($\omega_y$). (Student's \emph{t}-distribution) 95\% confidence intervals~\cite{Riley2006} around the separation data are shown in light grey.  Predictions using Eq.~\eqref{eq:screw_model_separation} with molecular pitch values are shown as straight lines. Red indicates the non-interfering solvent model, and blue is for interfering solvent.}
    \label{fig:sepvsshear}
\end{figure}

The average separation of enantiomers in the $y$ direction was measured using the mean displacement correlation function,
\begin{equation}
    \langle \delta y(t) \rangle = \langle y_i(t+\tau) - y_i(\tau) \rangle_{i,\tau}
\label{eq: meandisp}
\end{equation}
for enantiomers $i$ in one of the two RNEMD regions. The correlations were averaged over initial times $\tau$ and over all the enantiomers present in the RNEMD region. The final separation result (shown as black curves in Fig. \ref{fig:sepvsshear}) was computed from the difference in displacements (Eq. (\ref{eq: meandisp})) of the ($R$) and ($S$) enantiomers, and was averaged over the number of sampling regions (two for each simulation). Predictions of separation were done using a simple competition model previously outlined by Duraes and Gezelter,\cite{Drisko2024} which predicts the linear drift velocities of the solution and enantiomers from their mole fractions, the local vorticity, and the molecular pitch. The separation between enantiomers, $d(t)$, is directly related to the scalar pitch $(|P|/2\pi)$ of the enantiomers and the solution vorticity $(\omega_y)$,
\begin{equation}
\label{eq:screw_model_separation}
d(t)= \left| \left< \delta y_R(t)\right> - \left<\delta y_S(t)\right> \right| = \frac{|P|}{2\pi} \cdot \omega_y \pi \, t \cdot (x_R + x_S)
\end{equation}
where $x_R=\displaystyle\frac{n_{R}}{n_{R}+n_{S}+n_\mathrm{solvent}}$ is the mole fraction of the ($R$) enantiomers and $x_S$ is the mole fraction of the ($S$) enantiomers. From this model, two important limits emerge depending on the interference of solvent in enantiomeric displacement. The \textit{non-interfering solvent} limit assumes that the motion of enantiomers depends only on the concentrations of the two enantiomers (e.g., $x_R \rightarrow \left(\frac{n_{R}}{n_{R}+n_{S}}\right)$), while the \textit{interfering solvent} model takes into account the driving force from one solute pushing on the solvent molecules surrounding the other solutes. In Fig. \ref{fig:sepvsshear}, both models are represented by red (non-interfering) and blue (interfering) curves. We note that the separation curve for the free acid form of montelukast was obtained from three different sets of simulations (each with 5 replicas) using multiple values of the momentum flux; the details of which can be found in SI. 

We note that some solute and solvent pairings could exhibit a time dependence of the scalar pitch that would impact the linear separation approximation in Eq. \eqref{eq:screw_model_separation}. In the SI, we provide time correlations of pitch values to help understand the dynamics of fluctuations around the mean, for bicalutamide in acetone and for montelukast sodium in ethanol. We observe a rapid decay of the scalar pitch ($50-100$ ps), indicating that there is only a short memory in this solute / solvent pairing. Over a 10~ns timescale, it is therefore reasonable to utilize the mean pitch approximation.

Both bicalutamide and montelukast sodium fall between the two limits of the competition models. Over a period of 2 ns under continuous shear, the enantiomers separated at a linear rate comparable to results obtained by Duraes and Gezelter.\cite{Duraes2021} With larger molecular pitch values than those used in previous rigid-body simulations, we expected bicalutamide and montelukast to achieve even better separation. However, this was only the case for bicalutamide, whose enantiomeric separation in acetone closely follows the non-interfering solvent model. Despite having a higher scalar pitch, the separation of montelukast sodium and its free acid was hindered by solvent interference. The interfering solvent model suggests that a translating montelukast enantiomer will push solvent molecules in one direction, opposing counter-translation of the other enantiomer. Ethanol is also more viscous than acetone, limiting the use of higher shear stress to reduce frictional heating and potentially preventing a higher degree of achievable separation.       

These results suggest that although ethanol is required to overcome solubility issues of montelukast sodium, it is a less ideal solvent for shear flow separation. Indeed, solubility is a primary concern in the success of this separation method. Initial attempts to simulate the separation of bicalutamide and montelukast in SPC/E water were unsuccessful due to aggregation of the solute molecules, while organic solvents such as ethanol and acetone reduced the likelihood of aggregation. Regions containing aggregations of enantiomers tend to exhibit local depression of vorticity, which affects bulk vorticity and reduces rotational motions of embedded enantiomers. The ionic form (montelukast sodium) showed better solubility in ethanol; however, the presence of charged counter ions had significant impacts on the independent flow of the enantiomers, producing a non-linear separation curve. We expect that solvent choice and solute concentration will be essential criteria for evaluating effective shear flow separation, with emphasis on higher solubility and lower concentrations of the enantiomers. 

To determine if separation is possible within a reasonable time frame, we utilize the coupled drift-diffusion model for enantiomers ($R$) and ($S$),\cite{Duraes2021} 
\begin{equation}
\renewcommand{\arraystretch}{0.8}
\label{eq:coupled_drift_diffusion}
\frac{\partial}{\partial t}\left(\begin{array}{c}
c_{R}\\
c_{S}
\end{array}\right)=D_\mathrm{tt}\,\frac{\partial^{2}}{\partial y^{2}}\left(\begin{array}{c}
c_{R}\\
c_{S}
\end{array}\right)-\frac{|P| \omega_{y}}{2 \rho_{o}}\left(\begin{array}{cc}
2 & -1\\
1 & -2
\end{array}\right)\left(\begin{array}{c}
c_{R}\\
c_{S}
\end{array}\right)\frac{\partial}{\partial y}\left(\begin{array}{c}
c_{R}\\
c_{S}
\end{array}\right)
\end{equation}
where $c_{R}=c_{R}\left(y,t\right)$, $c_{S}=c_{S}\left(y,t\right)$ are the concentrations of ($R$) and ($S$) enantiomers as a function of the $y$-coordinate and the time $t$, and $D_\mathrm{tt}$ is the molecule-specific translational diffusion coefficient, which can be computed from the trace of its translational diffusion tensor. Translational diffusion can overwhelm separation due to shearing, so the ratio between shear-induced separation and translational diffusion,\cite{Duraes2021}
\begin{equation}
s = \frac{|P| \omega_y}{2 \rho_o D_\mathrm{tt}}, 
\label{eq:ratio} 
\end{equation} 
is a useful test of whether shear separation is possible.

We note that Eq. \eqref{eq:coupled_drift_diffusion} treats the enantiomers in the limit of low concentrations, where aggregation between the enantiomeric molecules is disfavored. A more complete model would treat the aggregation and disaggregation of the enantiomers as a series of reversible reactions leading to aggregates of increasing mass. This is the long-standing coagulation-fragmentation problem, and the kinetics of this process have been explored in a mean field treatment.\cite{WATTIS20061}  In the regime where aggregation kinetics dominate over shear-induced drift, the concentrations of the two isolated enantiomers would drop to zero everywhere, and shear separation would be ineffective.

Duraes and Gezelter carried out drift-diffusion simulations starting from initial concentration profiles that were Gaussian, $c_{S,R}\left(y, 0\right)= c_0 ~e^{-\pi y^{2}}$, injected around $y=0$ and $t=0$.\cite{Duraes2021} We note that in order to observe separation in this model, non-uniform concentration profiles are required, given the gradient terms in Eq. \eqref{eq:coupled_drift_diffusion}. 
They found that racemic mixtures eventually separated on a centimeter length scale as long as this ratio exceeded $1~\mathrm{L~mol^{-1}~cm^{-1}}$.\cite{Duraes2021} We computed this ratio for our simulations of bicalutamide and montelukast sodium to be $1.23 \times 10^5$ and $1.14 \times 10^5$ $\mathrm{L~mol^{-1}~cm^{-1}}$, respectively. The simulated linear separation rates in Fig. \ref{fig:sepvsshear} suggest that successful enantiomeric separation of these targets can be achieved, provided that the imposed solution vorticity is large enough to overcome translational diffusion. However, this is not always the case for practical separation devices, and maintaining consistent vorticity also presents a formidable challenge. In the next section, we explore a proposed design for a device capable of enantiomeric separation through constant local shear vorticity.

\subsection{Taylor-Couette devices for enantiomeric separation}

Our simulation conditions and results suggest that it will be possible to carry out an experimental demonstration of enantiomeric separation using shear vorticity. There have been many attempts to generate shear vorticity for the purposes of chiral resolution ever since Howard's rotating drum experiment.\cite{Howard1976,Hirschfelder1977} Specifically, Hermans \textit{et al.} proposed a separation method based on the use of the Taylor-Couette flow.\cite{Hermans2015,Hermans2019}

\begin{figure}[H]
    \centering
    \includegraphics[width=0.9\textwidth]{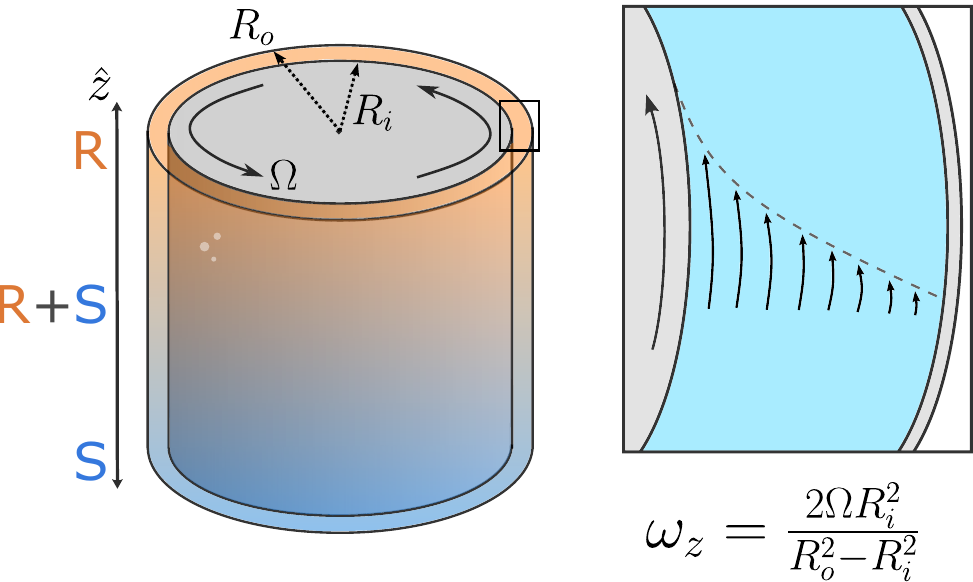}
    \caption{Left: A simple Taylor-Couette device which applies constant vorticity to a solution in order to separate a racemic mixture of solute molecules. A thin fluid film separates a fixed outer cylinder ($R_o$) from an inner cylinder ($R_i$) that is rotating with angular velocity $\Omega$. Right: below the critical Taylor number for the fluid, this creates a region of laminar shear, which has a constant vorticity ($\omega_z$) in the axis parallel to the cylinders.}
    \label{fig:taylordevice}
\end{figure}

A Taylor-Couette flow device consists of two concentric cylinders separated by a thin layer of viscous fluid (see Fig. \ref{fig:taylordevice}). The viscous fluid has a constant density $\rho$ and a constant dynamic viscosity $\mu$. Given that there is no axial movement in the $z$ direction, rotating the inner cylinder at an angular velocity $\Omega$ while keeping the outer cylinder fixed creates a steady, laminar, and azimuthal shear flow known as Couette flow. This type of flow has been extensively studied going back to a foundational work by Couette and explained succinctly by Lamb.\cite{Couette1890,Lamb1945} By invoking stick boundary conditions, the analytical form of the laminar, azimuthal velocity profile can be expressed as, 
\begin{equation}
    \nu_{\theta}(R) = \left(\frac{{R_i}^2 ~ \Omega}{R_o^2-R_i^2}\right)R - \left(\frac{R_i^2 ~ \Omega ~ R_o^2}{R_o^2-R_i^2}\right)\frac{1}{R} 
\label{eq: couettevelocity}
\end{equation}
where $R_o$ and $R_i$ are the radii of the outer and inner cylinders, respectively, and $\Omega$ is the angular velocity of the inner cylinder. Applying the vorticity equation $\omega = \nabla\times\nu$ in cylindrical coordinates to Eq. (\ref{eq: couettevelocity}) allows us to compute the axial vorticity inside the thin fluid layer,
\begin{equation}
    \omega_z = - \frac{2 ~ \Omega ~ R_i^2 }{R_o^2-R_i^2}
\label{eq:couettevorticity}
\end{equation}
which points in the vertical $(z)$ direction. Eq. (\ref{eq:couettevorticity}) suggests that the ($R$) and ($S$) enantiomers can be collected at the two vertical ends of the cylindrical system if a racemic mixture is injected at the midpoint. However, a major drawback of the Couette flow is the instability of the flow as the rate of rotation increases, which limits the achievable maximum vorticity before turbulence sets in. The phenomenon was studied by Taylor,\cite{Taylor1923} whose dimensionless stability criterion was later termed the \textit{Taylor number}. For a circular Couette flow, the Taylor number has a critical limit,\cite{Taylor1923}
\begin{equation}
\mathrm{Ta}_\mathrm{crit} = \frac{R_i~(R_i-R_o)^3 ~ \Omega^2}{\nu^2} ~ \approx ~ 1700
\label{eq: critTa}
\end{equation}
where $\nu$ is the kinematic viscosity of the fluid. Rotating the inner cylinder too fast will exceed this limit and will introduce vertical Taylor vortices that can complicate and oppose separation efforts. Eq. (\ref{eq: critTa}) allows us to compute the upper limit for the vorticity associated with different angular velocity $\Omega$ and inner cylinder radii $R_i$ values, assuming the thin fluid layer is kept constant at 0.1cm. Combining these vorticity values with simulation parameters, \textit{e.g.}, solution density, translational diffusion constant, and solution vorticity, as well as pitch information, the separation ratio $s$ predicts the feasibility of enantiomeric separation using a Taylor-Couette device.  For values of $s > 1$, separation is possible in this device, and for $s < 1$, separation would not be achievable.
Using the upper Taylor number limit and our simulation conditions, we have predicted the shear vorticity for various device parameters. This allows us to compute the separation-diffusion ratio from Eq. (\ref{eq:ratio}) to assess the practicality of this device, as shown in Table. \ref{tab:taylorcouette}. 

\begin{table}[H]
\renewcommand{\arraystretch}{1.25}
\caption{Taylor-Couette device parameters necessary for separation of enantiomers. This device comprises two concentric cylinders with a 1 mm liquid film between the outer (fixed) cylinder ($R_o = R_i + 0.1$) and the inner cylinder. The inner cylinder rotates with angular velocity $\Omega$ and produces an average vorticity $\bar{\omega}$ in the fluid film. All the $\Omega$ and $\langle \omega \rangle$ values listed are upper limits calculated from the critical Taylor number. The resulting separation ratio, $s$ (see Eq. \eqref{eq:ratio}), must be greater than 1.0 to effectively separate the enantiomers.} \label{tab:taylorcouette}
\begin{tabular}{|c|rrc|rrc|}
    \hline
    &\multicolumn{3}{c}{Bicalutamide in acetone} & \multicolumn{3}{|c|}{Montelukast sodium in ethanol} \\
    \hline
    $R_i$ (cm) & $\Omega$ (s$^{-1}$) & $\langle \omega \rangle$ (s$^{-1}$) & $s$ (L mol$^{-1}$ cm$^{-1}$) & $\Omega$ (s$^{-1}$) & $\langle \omega \rangle$ (s$^{-1}$) & $s$ (L mol$^{-1}$ cm$^{-1}$) \\
    1   & 87.3& $5.23 \times 10^3$ & 0.04 & 286.7& $1.72 \times 10^4$ & 0.58 \\
    10  & 27.6& $1.73 \times 10^4$ & 0.15 & 90.7 & $5.67 \times 10^4$ & 1.93 \\
    20  & 19.5& $2.45 \times 10^4$ & 0.21 & 64.1 & $8.04 \times 10^4$ & 2.73 \\
    50  & 12.4& $3.88 \times 10^4$ & 0.33 & 40.5 & $1.27 \times 10^5$ & 4.33 \\
    100 & 8.7 & $5.49 \times 10^4$ & 0.46 & 28.7 & $1.80 \times 10^5$ & 6.12 \\
    200 & 6.2 & $7.76 \times 10^4$ & 0.65 & 20.3 & $2.55 \times 10^5$ & 8.66 \\
    500 & 3.9 & $1.23 \times 10^5$ & 1.03 & 12.8 & $4.03 \times 10^5$ & 13.69\\
    1000& 2.8 & $1.74 \times 10^5$ & 1.46 & 9.1  & $5.70 \times 10^5$ & 19.36\\
    \hline \hline
    
\end{tabular}
\end{table}
    
From these calculations, it appears to be impossible to separate bicalutamide enantiomers using a Taylor-Couette device unless the inner cylinder has a radius larger than 5~m. This is due to: 1) low vorticity generated from a lower angular velocity limit imposed by the Taylor's number criterion, and 2) the low viscosity of acetone as the solvent. The more viscous ethanol-based solutions for montelukast allow the device to work at a much smaller inner radius ($>10$ cm), which is in the size range of commercial viscometers. A meter-sized inner cylinder can bring the shear vorticity up to values comparable to our simulated values, although the fluid film needs to stay sufficiently thin, as a wider gap increases the Taylor number of the flow regime. With molecular pitch as the theoretical basis for enantiomeric separation, it therefore appears possible to use Taylor-Couette devices to perform these separations, although high solvent viscosity becomes one of the main factors for feasibility. Further considerations include: injection and collection methods, the height of the device, the means of rotation, as well as local friction and dissipation of heat in the shearing fluid.  

\section{Conclusion}
We have studied the effects of molecular flexibility on the molecular pitch model and on the possibility of enantiomeric separation using shear flow.  The two targets of this study, bicalutamide and montelukast sodium, are pharmaceutical molecules with large inherent molecular pitch values in their optimized (gas-phase) geometries. Simulations of a single enantiomers in various solvents confirmed a flexibility-induced distribution of molecular pitch as a result of conformational changes. However, these distributions are not highly dependent on solvent identity. Compared with optimized gas phase geometries, molecular flexibility and conformational freedom produced a lower mean scalar pitch in montelukast sodium and a higher scalar pitch in bicalutamide. This is likely caused by their varying degrees of flexibility. Montelukast sodium is more flexible making it more difficult for the molecule to retain high-pitch conformations in solution.  We also found that solution shear did not produce observable changes to the pitch distribution of bicalutamide while slightly increasing the probability of lower-pitch conformations in montelukast sodium. 

Under continuous shear, a racemic solution of bicalutamide enantiomers exhibited a linear separation rate that was much larger than that of montelukast sodium. Separation on molecular dynamics time scales is limited by solvent viscosity, and by aggregation of solute molecules. However, this may not limit the applicability of this technique on experimental time scales. In the case of shear flow devices, enantiomeric separation via solvent shear is in direct competition with translational diffusion, so a high solvent viscosity may make realistic separations possible while simultaneously making MD simulations nearly intractable. Although high solution vorticities were imposed in our molecular simulations, both molecules showed potential for experimental separation, as predicted by a ratio $s$ (see Eq. \eqref{eq:ratio}) between separation drift and diffusive mixing. Given a linear separation curve at short times and $s>1$, enantiomeric separation on a centimeter-length scale should be possible with a few hours under shear-induced vorticity. 

We also predict that Taylor-Couette devices can create stable shear flows for enantiomeric separations. These devices will exhibit flow instability when the Taylor number exceeds a critical value ($\mathrm{Ta} > 1700$), effectively limiting the maximum rate of rotation. Thus, the low viscosity of acetone makes it impossible for shear flow separation inside this device to overcome translational diffusion unless either unphysical rotational rates or large cylinder radii are involved.  Our results suggest that practical shear flow separation is highly dependent on the details of the solvent and solute molecules. Low viscosity solvents are likely unusable for this process, as are separation targets that easily aggregate. For prospective shear flow separation using Taylor-Couette devices, we therefore recommend using a high viscosity solvent that easily dissolves the molecules to be separated.

\section{Data and Software Availability}
The data used in this study (force field parameters, initial configurations, extracted data, graphs, scripts, and meta-data) are available at DOI: \href{https://doi.org/10.5281/zenodo.19206242}{10.5281/zenodo.19206242}  All simulations utilized the OpenMD molecular dynamics engine (See \href{https://github.com/OpenMD/OpenMD}{github.com/OpenMD/OpenMD}),\cite{Drisko2024} which is available under a BSD 3-clause license . All analysis tools utilized here have been built into OpenMD. 

\begin{acknowledgement}
Support for this project was provided by the National Science Foundation under grant CHE-1954648. Computational time was provided by the Center for Research Computing (CRC) at the University of Notre Dame.
\end{acknowledgement}

\begin{suppinfo}
Force field parameters, particularly atom type properties for all molecules and solvents. System construction and simulation protocols. Additional figures for discussions of shear-induced rotational motion, phase-separation and aggregation, and pitch fluctuation autocorrelation functions. Separation data from simulations utilizing other solution vorticities.
\end{suppinfo}

\section{Author Contributions}
This work was made available through contributions from  all  authors.  All  authors have approved the final version of the manuscript. M.N.P. and J.D.G. conceived and designed the simulations; M.N.P.  and L.C.  performed the simulations; M.N.P. and J.D.G. analyzed the data; M.N.P. and J.D.G. wrote and edited the manuscript. J.D.G. secured financial support for the research.

\section{Notes}
The authors declare no competing financial interest.

\newpage

\bibliography{sources}
\end{document}


\section{Force Field Parameters}

Force field parameters for all molecular dynamic (MD) simulations were adapted from the Generalized Amber Force Field version 2 (GAFF2).\cite{Wang2004} All atoms were given GAFF2 atom types, and their molecular charges were obtained from several sources. Partial charges for bicalutamide, montelukast sodium (and its free acid form), benzene, and acetone were computed using the AM1-BCC semi-empirical charge model.\cite{Jakalian2002,Silva2012,Kagami2023} The work of Fennell, Wymer, and Mobley provided partial charges for the ethanol and methanol models, which ensured an accurate description of their solvation free energies.\cite{Fennell2006} Water was modeled using the rigid SPC/E water model from  Berendsen, \textit{et al.},\cite{Berendsen1987} while the monoatomic sodium cation was modeled using values specified by Li, Song, and Merz.\cite{Li2015} All simulations utilized the damped shifted force method for  electrostatic interactions with a damping parameter ($\alpha$) of 0.185 \AA$^{-1}$ and a real space cut-off radius of 12 \AA.\cite{Fennell2006}  Geometries for ($S$)-bicalutamide and ($S$)-montelukast (anion and free acid) were optimized in OpenMD using steepest descent (SD).

\begin{figure}[H]
    \centering
    \includegraphics[width=0.8\textwidth]{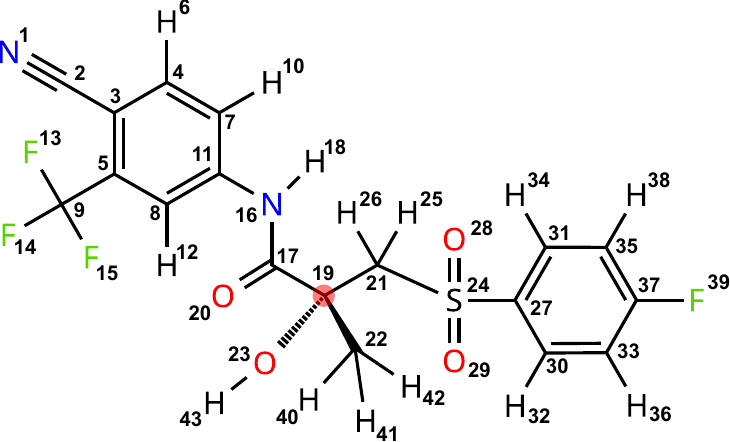}
    \caption{Two-dimensional structure of ($R$)-bicalutamide with atomic indices used in simulations. The chiral center is highlighted in red.}
    \label{fig:casodex structure SI}
\end{figure}

\begin{center}
\small
\renewcommand{\arraystretch}{0.9}
\begin{longtable}{c|ccr}
    \caption{GAFF2 atom types and partial charges of bicalutamide listed by atomic indices.}
    \label{tab:casodex inc} \\
    
        \hline \multicolumn{1}{|c|}{Index} & Element & GAFF2 Atom Type & \multicolumn{1}{c|}{Charge} \\ \hline
    \endfirsthead
        \hline \multicolumn{1}{|c|}{Index} & Element & GAFF2 Atom Type & \multicolumn{1}{c|}{Charge} \\ \hline
    \endhead
        \hline \multicolumn{4}{|r|}{{Continued on next page}} \\ \hline
    \endfoot
        \hline \hline
    \endlastfoot

    1  & N & n1 & -0.328800 \\
    2  & C & cg & 0.219800 \\
    3  & C & ca & 0.002000 \\
    4  & C & ca & -0.067000 \\
    5  & C & ca & -0.099300 \\
    6  & H & ha & 0.155000 \\
    7  & C & ca & -0.173000 \\
    8  & C & ca & -0.128000 \\
    9  & C & c3 & 0.686200 \\
    10 & H & ha & 0.143000 \\
    11 & C & ca & 0.096600 \\
    12 & H & ha & 0.200000 \\
    13 & F & f  & -0.227967 \\
    14 & F & f  & -0.227967 \\
    15 & F & f  & -0.227967 \\
    16 & N & ns & -0.476100 \\
    17 & C & c  & 0.663100 \\
    18 & H & hn & 0.320500 \\
    19 & C & c3 & 0.144800 \\
    20 & O & o  & -0.532100 \\
    21 & C & c3 & -0.335600 \\
    22 & C & c3 & -0.143100 \\
    23 & O & oh & -0.588800 \\
    24 & S & sy & 1.394301 \\
    25 & H & h1 & 0.120200 \\
    26 & H & h1 & 0.120200 \\
    27 & C & ca & -0.430500 \\
    28 & O & o  & -0.658300 \\
    29 & O & o  & -0.658300 \\
    30 & C & ca & 0.015500 \\
    31 & C & ca & 0.015500 \\
    32 & H & ha & 0.171500 \\
    33 & C & ca & -0.196000 \\ 
    34 & H & ha & 0.171500 \\
    35 & C & ca & -0.196000 \\
    36 & H & ha & 0.164000 \\
    37 & C & ca & 0.203900 \\
    38 & H & ha & 0.164000 \\
    39 & F & f  & -0.122900 \\
    40 & H & hc & 0.066700 \\
    41 & H & hc & 0.066700 \\
    42 & H & hc & 0.066700 \\
    43 & H & ho & 0.446000 \\
\end{longtable}
\end{center}

\begin{figure}[H]
    \centering
    \includegraphics[width=\textwidth]{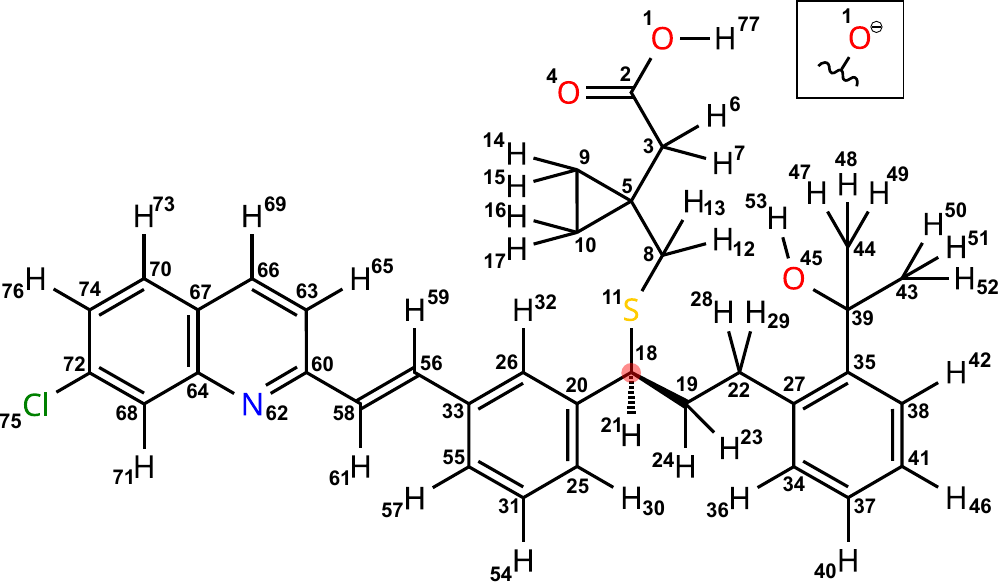}
    \caption{Two-dimensional structures of ($R$)-montelukast (free acid) and its anionic form (in square box) with atomic indices used in simulations. The chiral center is highlighted in red.}
    \label{fig:singulair structure SI}
\end{figure}

\begin{center}
\small
\renewcommand{\arraystretch}{0.9}
\begin{longtable}{c|ccrr}
    \caption{GAFF2 atom types and partial charges of montelukast (anion and free acid) listed by atomic indices. The carboxylic O is listed with two different GAFF2 atom types depending on the form of montelukast.}
    \label{tab:singulair inc} \\
    
        \hline \multicolumn{1}{|c|}{Index} & Element & GAFF2 Atom Type & Charge (anion) & \multicolumn{1}{c|}{Charge (free acid)} \\ \hline
    \endfirsthead
        \hline \multicolumn{1}{|c|}{Index} & Element & GAFF2 Atom Type & Charge (anion) & \multicolumn{1}{c|}{Charge (free acid)} \\ \hline
    \endhead
        \hline \multicolumn{5}{|r|}{{Continued on next page}} \\ \hline
    \endfoot
        \hline \hline
    \endlastfoot
    
    1  & O & o (Anionic) & -0.851800 & -0.581100 \\
       &   & oh (Free acid) &  & \\
    2  & C & c  & 0.909602 & 0.627100 \\
    3  & C & c3 & -0.175400 & -0.147400 \\
    4  & O & o  & -0.851800 & -0.495000 \\
    5  & C & cx & -0.053000 & -0.124000 \\
    6  & H & hc & 0.037700 & 0.094200 \\
    7  & H & cc & 0.037700 & 0.094200 \\
    8  & C & c3 & 0.038700 & 0.011700 \\
    9  & C & cx & -0.137400 & -0.119400 \\
    10 & C & cx & -0.137400 & -0.119400 \\
    11 & S & ss & -0.353200 & -0.315200 \\
    12 & H & h1 & 0.078700 & 0.071700 \\
    13 & H & h1 & 0.078700 & 0.071700 \\
    14 & H & hc & 0.063200 & 0.080950 \\
    15 & H & hc & 0.063200 & 0.080950 \\
    16 & H & hc & 0.063200 & 0.080950 \\
    17 & H & hc & 0.063200 & 0.080950 \\
    18 & C & c3 & 0.077700 & 0.069700 \\
    19 & C & c3 & -0.054400 & -0.062400 \\
    20 & C & ca & -0.094300 & -0.102300 \\
    21 & H & h1 & 0.054700 & 0.068700 \\
    22 & C & c3 & -0.062100 & -0.049100 \\
    23 & H & hc & 0.042700 & 0.060200 \\
    24 & H & hc & 0.042700 & 0.060200 \\
    25 & C & ca & -0.089000 & -0.109000 \\
    26 & C & ca & -0.104000 & -0.110000 \\
    27 & C & ca & -0.054300 & -0.070300 \\
    28 & H & hc & 0.080700 & 0.062200 \\
    29 & H & hc & 0.080700 & 0.062200 \\
    30 & H & ha & 0.223000 & 0.154000 \\
    31 & C & ca & -0.130000 & -0.126000 \\
    32 & H & ha & 0.122000 & 0.131000 \\
    33 & C & ca & -0.110800 & -0.072800 \\ 
    34 & C & ca & -0.145000 & -0.138000 \\
    35 & C & ca & -0.052300 & -0.066300 \\
    36 & H & ha & 0.124000 & 0.130000 \\
    37 & C & ca & -0.141000 & -0.126000 \\
    38 & C & ca & -0.110000 & -0.103000 \\
    39 & C & c3 & 0.200100 & 0.203100 \\
    40 & H & ha & 0.116000 & 0.131000 \\
    41 & C & ca & -0.152000 & -0.139000 \\
    42 & H & ha & 0.150000 & 0.156000 \\
    43 & C & c3 & -0.135600 & -0.124100 \\
    44 & C & c3 & -0.135600 & -0.124100 \\ 
    45 & O & oh & -0.627800 & -0.607800  \\
    46 & H & ha & 0.117000 & 0.131000 \\
    47 & H & hc & 0.058533 & 0.047700 \\
    48 & H & hc & 0.058533 & 0.047700 \\
    49 & H & hc & 0.058533 & 0.047700 \\
    50 & H & hc & 0.058533 & 0.047700 \\
    51 & H & hc & 0.058533 & 0.047700 \\
    52 & H & hc & 0.058533 & 0.047700 \\
    53 & H & ho & 0.401000 & 0.401000 \\
    54 & H & ha & 0.152000 & 0.145000 \\
    55 & C & ca & -0.111000 & -0.105000 \\
    56 & C & ce & -0.012200 & -0.053200 \\
    57 & H & ha & 0.125000 & 0.140000 \\
    58 & C & cf & -0.295800 & -0.253800 \\
    59 & H & ha & 0.141000 & 0.151000 \\
    60 & C & ca & 0.505100 & 0.484100 \\
    61 & H & ha & 0.132000 & 0.127000 \\
    62 & N & nb & -0.674000 & -0.658000 \\
    63 & C & ca & -0.236300 & -0.239300 \\
    64 & C & ca & 0.416600 & 0.411600 \\
    65 & H & ha & 0.147000 & 0.143000 \\
    66 & C & ca & -0.086000 & -0.079000 \\
    67 & C & ca & -0.161300 & -0.153300 \\
    68 & C & ca & -0.162300 & -0.163300 \\
    69 & H & ha & 0.135000 & 0.140000 \\
    70 & C & ca & -0.098000 & -0.100000 \\
    71 & H & ha & 0.164000 & 0.165000 \\
    72 & C & ca & 0.021400 & 0.024400 \\
    73 & H & ha & 0.134000 & 0.139000 \\
    74 & C & ca & -0.136000 & -0.125000 \\
    75 & Cl & cl& -0.103400 & -0.088400 \\
    76 & H & ha & 0.144000 & 0.150000 \\
    77 & H & ho & & 0.429000 \\ 
    \end{longtable}
\end{center}

\begin{figure}[H]
    \centering
    \includegraphics[width=0.8\textwidth]{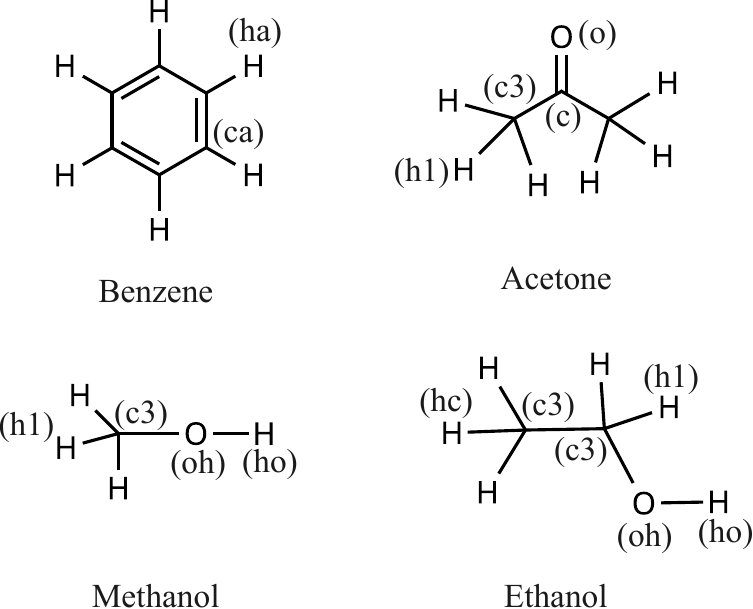}
    \caption{Two-dimensional structures of solvents (except SPC/E water) used in various simulations. GAFF2 atom types are assigned to a unique atom or set of atoms with identical partial charges.}
    \label{fig:solvent structure}
\end{figure}

\begin{center}
\small
\renewcommand{\arraystretch}{0.9}
\begin{longtable}{c|ccr}
    \caption{Unique GAFF2 atom types and partial charges of solvents used in various simulations.}
    \label{tab:solvent inc} \\

     \hline \multicolumn{1}{|c|}{Element} & GAFF2 Atom Type & Charge & \multicolumn{1}{c|}{Note} \\ \hline
    \endfirsthead
        \hline \multicolumn{1}{|c|}{Element} & GAFF2 Atom Type & Charge & \multicolumn{1}{c|}{Note} \\ \hline
    \endhead
        \hline \multicolumn{4}{|r|}{{Continued on next page}} \\ \hline
    \endfoot
        \hline \hline
    \endlastfoot

    \multicolumn{4}{l}{Acetone} \\ \hline
    C & c & 0.561102 &\\
    O & o & -0.531100 &\\
    C & c3 & -0.204100 &\\
    H & hc & 0.063033 &\\

    \hline \multicolumn{4}{l}{Benzene} \\ \hline
    H & ha & 0.13 &\\
    C & ca & -0.13 &\\

    \hline \multicolumn{4}{l}{Ethanol} \\ \hline
    H & hc & 0.0414000 &\\
    C & c3 & -0.1347150 & Bonded to hc\\
    C & c3 & 0.1556000 & Bonded to h1\\
    H & h1 & 0.0511400 &\\
    O & oh & -0.7269950 &\\
    H & ho & 0.4796300 &\\
    \hline \multicolumn{4}{l}{Methanol} \\ \hline
    C & c3 & 0.1452100 &\\
    O & oh & -0.7257900 &\\
    H & ho & 0.4801100 &\\
    H & h1 & 0.0334900 &\\

\end{longtable}
\end{center}

\begin{table}[H]
    \centering
    \renewcommand{\arraystretch}{1.5}
    \begin{tabular}{|c|ccccc|}
        \hline Element & GAFF2 Atom Type & $\epsilon$ (kcal/mol) & $\sigma$ (\AA) & Charge ($e$) & Sources \\ \hline
        Na & Na+ & 0.02909167 & 2.61033324415119 & 1.00 & Ref. \citen{Li2015} \\
        O & ow & 0.15532 & 3.16549 & -0.8476 & Ref. \citen{Berendsen1987} \\
        H & hw & & & 0.4238 & Ref. \citen{Berendsen1987} \\
        \hline \hline
    \end{tabular}
    \caption{GAFF2 atom types and Lennard-Jones parameters for sodium cation and rigid SPC/E water models. }
    \label{tab:Na SPCE}
\end{table}

\section{System Construction \& Simulation Protocol}
Initial configurations for the single-enantiomer simulations of montelukast sodium and bicalutamide in various solvents were generated using Packmol~\cite{Martinez2009} in orthorhombic boxes with periodic boundary conditions.\cite{Leach2001,Allen2017} The number of solvent molecules and box parameters were chosen so that the density of the simulation box matched the pure solvent at 298 K. To create five statistically independent solutions for each solvent, we resampled the initial velocities from a Maxwell-Boltzmann distribution at 298K. These simulation boxes were then equilibrated in the canonical (NVT) ensemble for 1 ns at 298.0 K using a Nose-Hoover thermostat,\cite{Hoover1985} after which they underwent 10-ns microcanonical (NVE) simulations. Data collection was carried out on these 10-ns simulations. The integration time step was set to 1 fs for all simulations.

This approach was also used to generate the initial configurations of racemic mixtures of bicalutamide and montelukast sodium in the appropriate solvents. Given their low aqueous solubility, ethanol and acetone were chosen as solvents for montelukast sodium and bicalutamide, respectively. The total number of enantiomers was set to 40 molecules (20 for each chirality) to yield effective solute concentrations of $0.3-0.4$~M.  See Table \ref{tab:box and RNEMD} for more details. This also helps prevent phase separation and solubility issues during equilibration. Solute molecules were evenly spaced along the $z$-axis of the simulation box to ensure a consistent density throughout. The equilibration phase consisted of a 0.5-ns canonical ensemble (NVT) simulation at 298.0 K, followed by another 0.5-ns microcanonical (NVE) simulation.

After equilibration, the simulation boxes were subjected to an imposed $x$-momentum flux along the $z$-axis of the box, $J_z(p_x)$ using the velocity shearing and scaling (VSS) reverse non-equilibrium molecular dynamics (RNEMD) algorithm in OpenMD.\cite{Kuang2012,Drisko2024} The flux is imposed between molecules residing in two separated regions of the box, each comprising 10\% of the box volume.  At steady state, molecules inside these `exchange regions' are propelled in opposing $x$-directions by the momentum exchange. The two regions between the exchange volumes which are known as `RNEMD regions', develop a linear velocity gradient in response to the imposed momentum flux.  Linear velocity gradients also characterize regions of uniform vorticity, $\mathbf{\omega} = \nabla \times \mathbf{v}$ in a viscous fluid. The resulting vorticity in each region  points in the opposing $y$- directions from the other. Rigid objects inside the RNEMD regions also acquire an angular velocity that is half of the vorticity $\omega_b = \omega / 2$.\cite{Duraes2021} Data collection was performed inside the RNEMD regions, including profiles of the local density, temperature, and velocity. Each simulation cell contains 2 counter-rotating RNEMD regions, so a total of 10 samples were averaged for the 5 RNEMD replicas.  With excessive imposed momentum flux, it is possible for frictional heating to occur inside the RNEMD region, so the magnitude of the momentum flux was chosen so that the local temperature differences would never exceed 10K over the length of the simulation cell.  It is also worth noting that these systems need a significant amount of time to reach steady-state velocity gradients, typically up to 1 ns into the RNEMD simulation.


\begin{table}[H]
\caption{Parameters for all RNEMD simulations. Cell properties are taken after the equilibration step. $n_\mathrm{X}$ denotes the number of molecules of type X in a given region. Concentrations are shown in square brackets with units of mol/L. Angle brackets, $\langle$ $\rangle$, are used for racemic mixture parameters that have been averaged over 10 samples and 5 ns of simulation time. $L_x, L_y, \mathrm{~and~} L_z$ give the dimensions of the relevant region.} 
\label{tab:box and RNEMD}
\renewcommand{\arraystretch}{1.3}
\begin{tabular}{c|lr|lr}
    \hline \multicolumn{1}{|c|}{Simulation} &\multicolumn{2}{c|}{Cell Properties} & \multicolumn{2}{c|}{In RNEMD Regions} \\
    \hline \multirow[t]{11}{6.0cm}{($R$)- and ($S$)-bicalutamide in acetone} & $n_\mathrm{S}$ & 20 & $\langle{n_\mathrm{S}}\rangle$ & 7.585 \\
    & $n_\mathrm{R}$ & 20 & $\langle{n_\mathrm{R}}\rangle$ & 8.154 \\
    & $n_{\mathrm{acetone}}$ & 1154 & $\langle{n_{\mathrm{acetone}}}\rangle$ & 461.697 \\
    & $[\mathrm{S}]$ & 0.184 & $\langle[\mathrm{S}]\rangle$ & 0.175 \\
    & $[\mathrm{R}]$ & 0.184 & $\langle[\mathrm{R}]\rangle$ & 0.188 \\
    & $[\mathrm{acetone}]$ & 10.644 & $\langle[\mathrm{acetone}]\rangle$ & 10.647 \\
    & $\langle{T}\rangle$ (K) & 298.32 & & \\
    & $L_x, L_y$ (\AA) & 30 & $L_x, L_y$ (\AA) & 30 \\
    & $L_Z$ (\AA) & 200 & $L_z$ (\AA) & 80 \\
    & $\tau_{zx}$ (mPa) & $4.51 \times 10^9$ & $\langle\eta\rangle$ (mPa s) & 0.33 \\
    & & & $\langle{|\omega_y|}\rangle$ ($\mathrm{s^{-1}}$) & $1.352 \times 10^{10}$ \\ 
    \hline \multirow[t]{11}{6.0cm}{($R$)- and ($S$)-montelukast sodium in ethanol} & $n_\mathrm{S}$ & 20 & $\langle{n_\mathrm{S}}\rangle$ & 7.557 \\
    & $n_\mathrm{R}$ & 20 & $\langle{n_\mathrm{R}}\rangle$ & 7.179 \\
    & $n_\mathrm{Na+}$& 40 & $\langle{n_\mathrm{Na+}}\rangle$& 14.485 \\
    & $n_{\mathrm{ethanol}}$ & 1804 & $\langle{n_{\mathrm{ethanol}}}\rangle$ & 674.851 \\
    & $[\mathrm{S}]$ & 0.151 & $\langle[\mathrm{S}]\rangle$ & 0.174 \\
    & $[\mathrm{R}]$ & 0.151 & $\langle[\mathrm{R}]\rangle$ & 0.166 \\
    & $[\mathrm{Na+}]$ & 0.302 & $\langle[\mathrm{Na+}]\rangle$ & 0.334 \\
    & $[\mathrm{ethanol}]$ & 13.637 & $\langle[\mathrm{ethanol}]\rangle$ & 12.754 \\
    & $\langle{T}\rangle$ (K) & 299.165 & & \\
    & $L_x, L_y$ (\AA) & 32.057 & $L_x, L_y$ (\AA) & 32.057 \\
    & $L_Z$ (\AA) & 213.716 & $L_z$ (\AA) & 85.487 \\
    & $\tau_{zx}$ (mPa) & $1.80 \times 10^{10}$ & $\langle\eta\rangle$ (mPa s) & 4.99 \\
    & & & $\langle{|\omega_y|}\rangle$ ($\mathrm{s^{-1}}$) & $3.618 \times 10^9$ \\ \hline \hline
\end{tabular}
\end{table}

\section{Shear-induced rotational motion}

In response to shearing-induced solution vorticity $(\omega)$, the solute molecules inside the RNEMD regions can rotate and acquire an angular velocity. The translation-rotation (tr) coupling block of the resistance tensor governs this process, and conversely, the (rt) block governs the translation resulting from molecular rotation. The molecular pitch theory predicts that the enantiomers will translate in opposing directions as a result of the same rotational motion.\cite{Duraes2023} To verify the generation of shear-induced rotation, to relate it to the solution vorticity, and to observe enantiomeric separation, we calculated the mean angular displacement, $\langle{cos\theta (t)}\rangle = \langle \hat{\mathbf{u}}(0) \cdot \hat{\mathbf{u}}(t) \rangle$, where $\hat{\mathbf{u}}(t)$ is the lab frame orientation of a molecule-fixed vector at time $t$. This is also known as the orientational correlation function. 

\begin{figure}[H]
    \centering
    \includegraphics[width=0.75\textwidth]{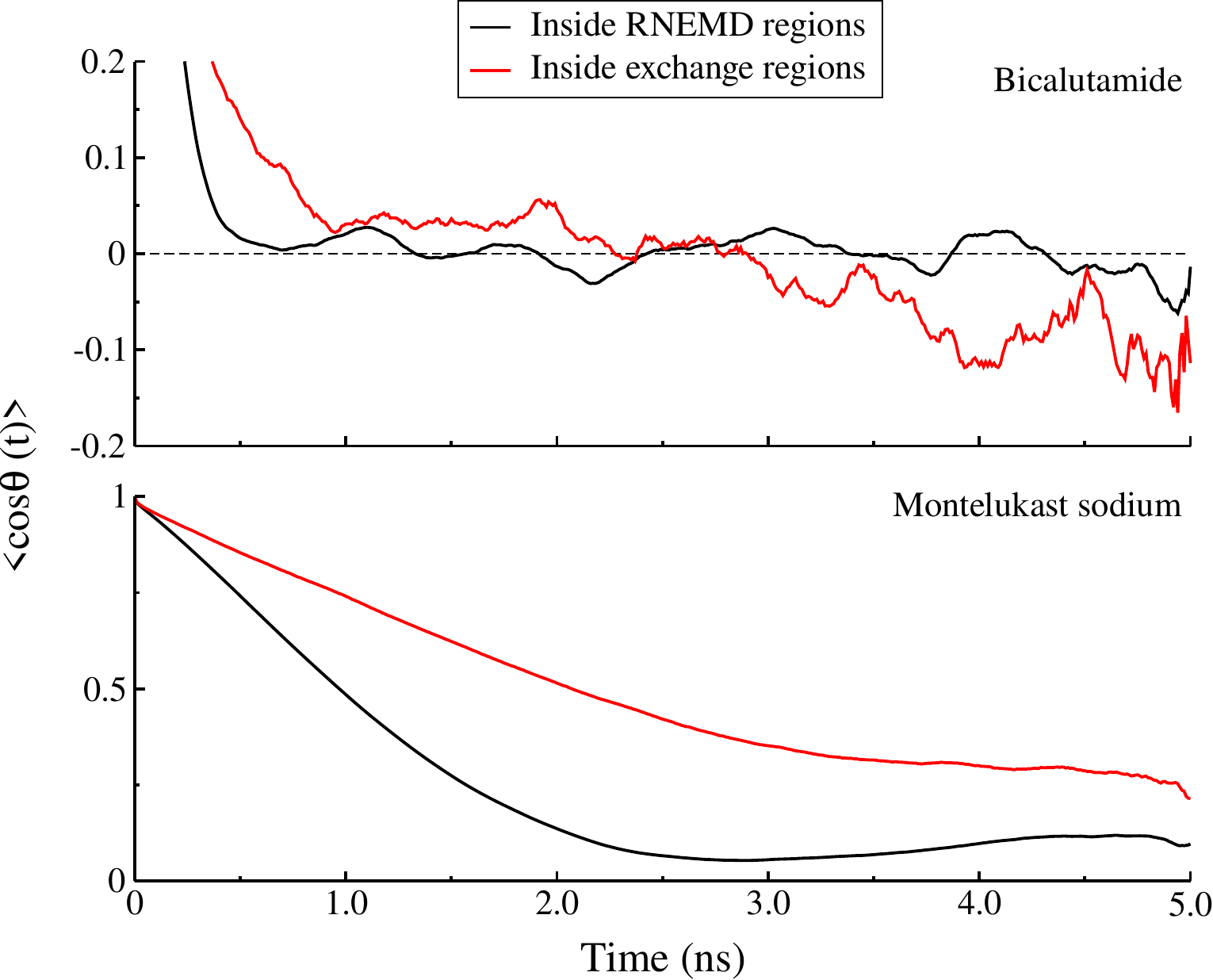}
    \caption{Orientational correlation functions of the enantiomeric molecules in this study. Black curves are computed for molecules in the RNEMD regions that are experiencing a uniform vorticity, while red curves represent those molecules in the exchange regions where the surrounding solvent is translating, but has negligible vorticity. In this correlation function, coherent rotational motion and tumbling will yield oscillatory or decaying behavior, respectively. Top: Bicalutamide in acetone. Bottom: Montelukast sodium in ethanol.}
    \label{fig:lcorr}
\end{figure}

The mean angular displacement measures the average orientation over time by correlating the unit vector pointed along a particular axis of the molecule between two times.  For bicalutamide, a vector was defined between the two N atoms belonging to one of the rigid `arms' of the molecule. In montelukast sodium, the vector spans the Cl atom and the aromatic C bonded to the chiral center (carbon 20), also along one of the rigid arms of the molecule. The curves in Fig. \ref{fig:lcorr} are averaged from both ($S$) and ($R$) enantiomers in all simulations with similar vorticities.  Due to tumbling and heterogeneous environments, the mean angular displacement decays over time from the value 1.  Oscillatory behavior indicates some degree of coherent rotational motion. 

Fig. \ref{fig:lcorr} suggests that inside the RNEMD regions, bicalutamide exhibits a greater degree of rotational motions in 5 nanoseconds than montelukast sodium. It rapidly decoheres and appears to oscillate, indicating a full molecular rotation on a 500 ps timescale. This also supported by the higher solution vorticity in bicalutamide / acetone simulations, which is associated with the low viscosity of acetone and the smaller size of bicalutamide. Montelukast sodium saw a much slower decay and fewer oscillations, indicating slower tumbling with fewer coherent rotational oscillations. This may be the cause of the non-linear enantiomeric separation curves for montelukast sodium (see Fig. 6 in the paper).

\section{Phase-separation and Aggregation}
One issue with solution-based simulations of these two drug molecules is their relatively low aqueous solubility. Attempts to use SPC/E water to solvate these molecules resulted in phase-separation and the formation of large aggregates even during equilibration. Therefore, for RNEMD simulations of racemic mixtures, we have utilized solvents in which each of these molecules was reportedly most soluble (shown in Table \ref{tab:box and RNEMD}). Despite these efforts, several instances where the enantiomers formed aggregates were still observed during all simulations. To better verify the existence of phase-separation, we have computed radial distribution functions for pairs of $(S)-(R)$ enantiomers and also between enantiomers and solvent atoms.

\begin{figure}
    \centering
    \includegraphics[width=\linewidth]{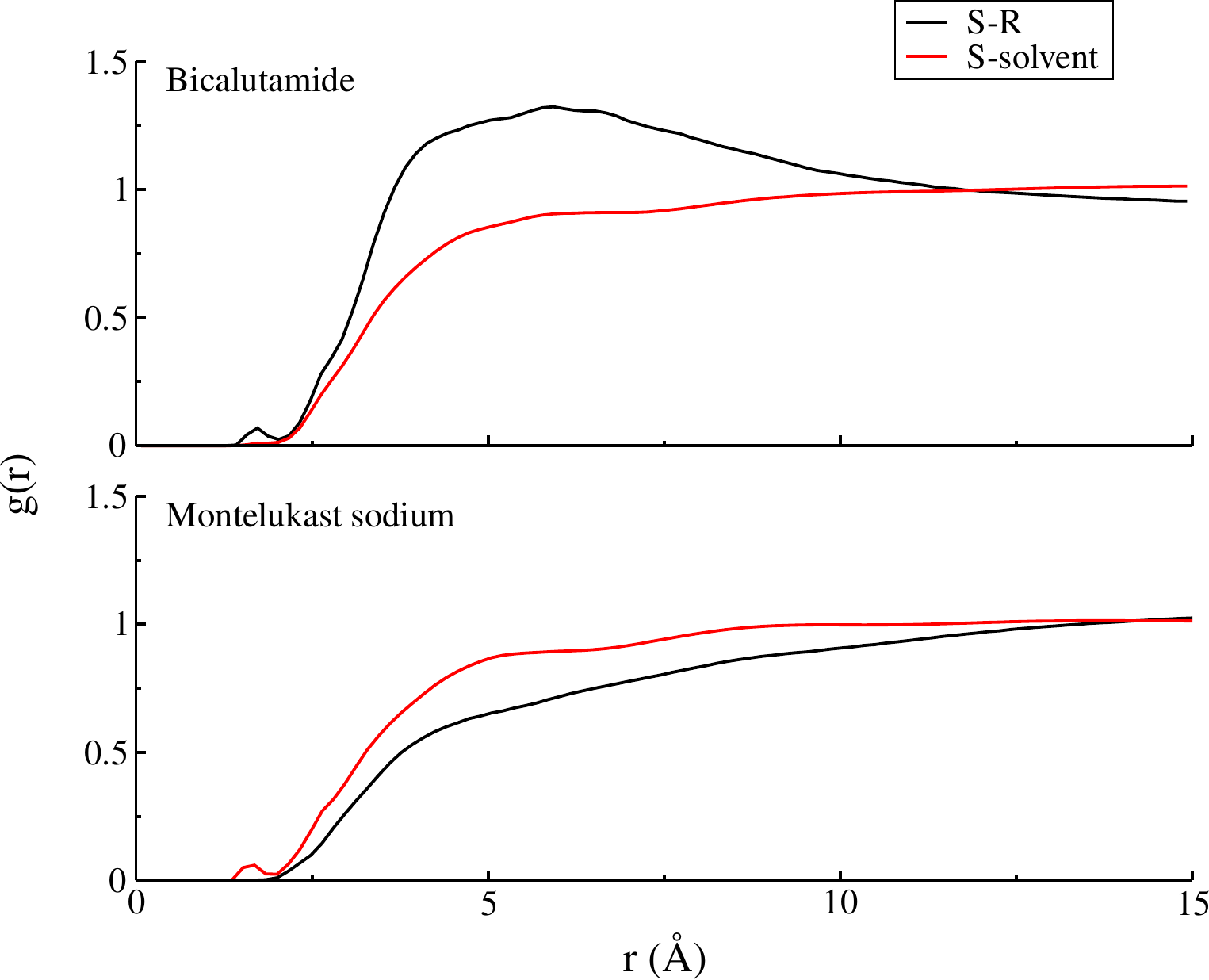}
    \caption{Radial distribution functions, $g_{S,R}(r)$ and $g_{S,\mathrm{solvent}}(r)$ for the racemic solutions of the two drug molecules in their respective solvents. Top: Bicalutamide in acetone. Bottom: Montelukast sodium in ethanol. We note that bicalutamide does appear to have a weak local preference for other bicalutamide molecules, while montelukast sodium has a closer attachment to the solvent.}
    \label{fig:gofr}
\end{figure}

The radial distribution profiles in Fig. \ref{fig:gofr} confirm our observations regarding the solubility issues of these racemic solutions. Bicalutamide appears to exhibit a higher degree of self-association between its enantiomers than montelukast sodium, while the enantiomers of the latter are better solvated by ethanol. At first glance, this appears to contradict the observed enantiomeric separation results, where bicalutamide enantiomers separate at a much higher rate under shear. Phase-separation should hinder enantiomeric separation by reducing coupling between the solute and solvent while simultaneously making aggregates harder to rotate and translate. 
Regions with a higher local density of enantiomers also exhibit a much lower effective solution vorticity. The higher viscosity of ethanol, however, may make it a more effective solvent for macroscopic separation methods explored in the main paper.

\section{Pitch autocorrelation and the validity of the mean pitch}

\begin{figure}
    \centering
    \includegraphics[width=\linewidth]{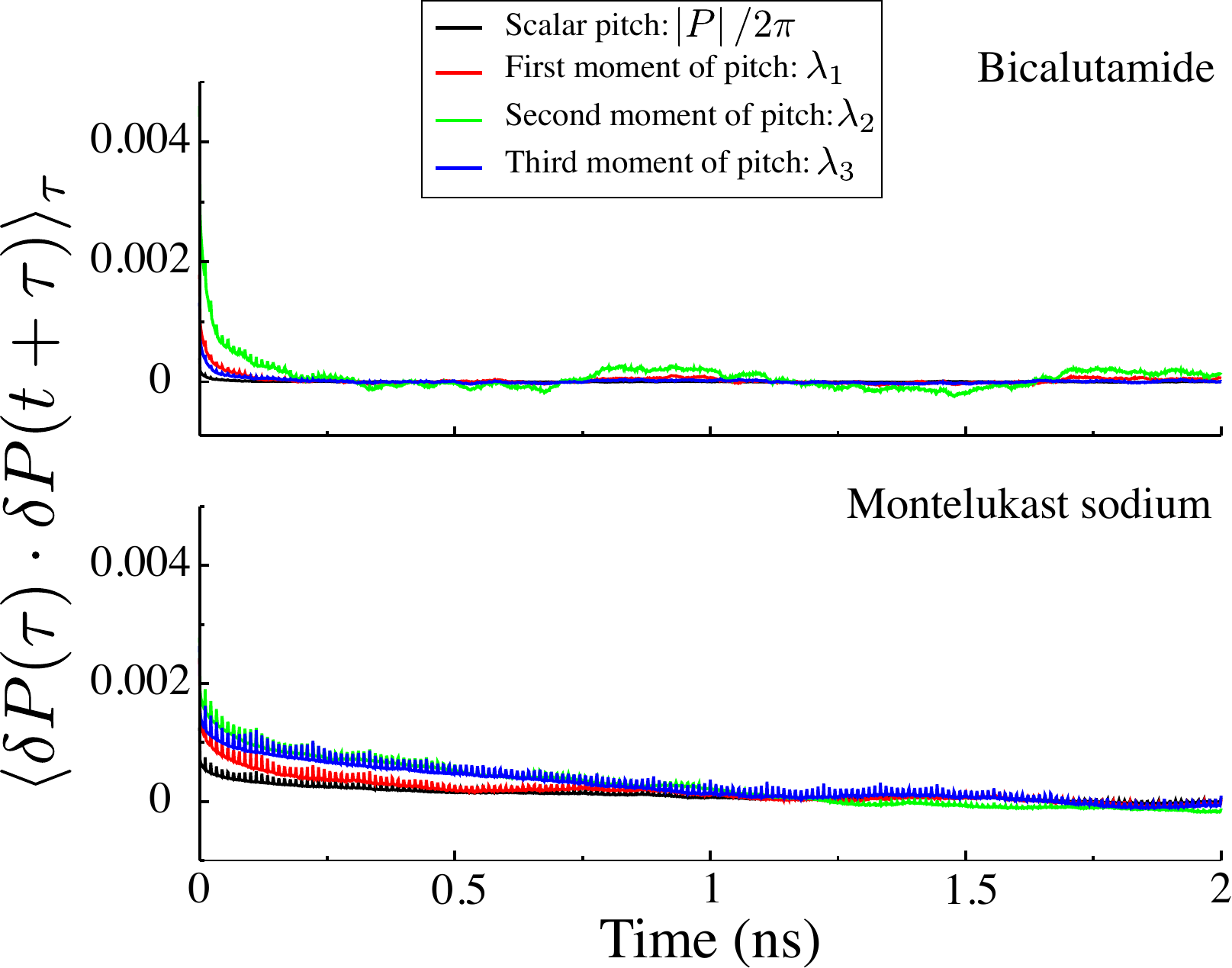}
    \caption{Pitch fluctuation autocorrelation functions for the two drug molecules in their respective solvents. Top: Bicalutamide in acetone. Bottom: Montelukast sodium in ethanol.}
    \label{fig:pitchACF}
\end{figure}

In the paper, we noted that some solute and solvent pairings could exhibit a time dependence of the scalar pitch that would impact the linear separation approximation in Eq. (8). Here, we provide time correlations of pitch fluctuations,
\begin{equation}
C_\text{pitch}(t) =  \langle \delta P(\tau) \cdot \delta P(\tau+t)\rangle_\tau
\end{equation}
where $\delta P(t) = P(t) - \langle P \rangle$. This autocorrelation function helps elucidate the dynamics of pitch fluctuations around the mean. In Fig. \ref{fig:pitchACF} we show these autocorrelation functions for the scalar pitch, $\left| P \right| / 2 \pi$ as well as the three moments of pitch $(\lambda_1, \lambda_2, \lambda_3)$ for bicalutamide in acetone and for montelukast sodium in ethanol. We observe a rapid decay of the scalar pitch, indicating that there is only a short memory in these solute / solvent pairings.  On a 10 ns timescale, it is therefore reasonable to utilize the mean pitch approximation.

\section{Effects of solution vorticity}

Given the non-linear enantiomeric separation of montelukast sodium, it is important to study whether the presence of charged molecules and ions have an effect on solution viscosity and separation output. We performed a set of RNEMD simulations of racemic solutions of the free acid form of montelukast while keeping the number of enantiomers and the solution density similar to  simulations of their ionic counterparts. Each set of simulations listed in Table \ref{tab:RNEMD free acid} was also carried out in the same manner and data were averaged from 5 statistically different simulations whose initial velocities were sampled from a Boltzmann distribution at 298~K. We also imposed three different values of the external momentum flux to evaluate how enantiomeric separation responds to changes in shear stress. 

\begin{table}[H]
\renewcommand{\arraystretch}{1.3}
\caption{Parameters for RNEMD simulations of racemic mixtures of free acid montelukast enantiomers with three different value of momentum flux $J_z (p_z)$ (equivalent to the shear stress $\tau_{zx}$). Cell properties are taken after the equilibration step. $n_\mathrm{X}$ denotes the number of molecules of type X in a given region. Concentrations are shown in square brackets with units of mol/L. Angle brackets, $\langle$ $\rangle$, are used for racemic mixtures’ parameters that have been averaged over 10 samples and 5 ns of simulation time. $L_x, L_y, \mathrm{~and~} L_z$ give the dimensions of the relevant region.}
\begin{tabular}{lr|lrrr}
    \hline \multicolumn{2}{|c|}{Cell Properties} & \multicolumn{4}{c|}{In RNEMD Regions} \\
    \hline \hline
    $n_\mathrm{S}$ & 20 & $\langle{n_\mathrm{S}}\rangle$ & 7.592 & 7.705 & 7.589 \\
    $n_\mathrm{R}$ & 20 & $\langle{n_\mathrm{R}}\rangle$ & 7.809 & 7.759 & 7.501 \\
    $n_\mathrm{ethanol}$ & 1681 & $\langle{n_\mathrm{ethanol}}\rangle$ & 676.429 & 675.722 & 677.884 \\
    $[\mathrm{S}]$ & 0.161 & $\langle{[\mathrm{S}]}\rangle$ & 0.152 & 0.154 & 0.152 \\
    $[\mathrm{R}]$ & 0.161 & $\langle{[\mathrm{R}]}\rangle$ & 0.156 & 0.155 & 0.150 \\
    $[\mathrm{ethanol}]$ & 13.551 & $\langle{[\mathrm{ethanol}]}\rangle$ & 13.551 & 13.537 & 13.580 \\
    $\langle{T}\rangle$ (K) & 298.17 & & & & \\
    $L_x, L_y$ (\AA) & 31.317 & $L_x, L_y$ (\AA) & 31.317 & 31.317 & 31.317 \\
    $L_Z$ (\AA) & 210 & $L_z$ (\AA) & 84.5 & 84.5 & 84.5 \\
    & & $\langle\eta\rangle$ (mPa s) & 5.81 & 4.51 & 3.98 \\
    & & $\langle{|\omega_y|}\rangle$ ($\mathrm{s^{-1}}$) & $1.552 \times 10^9$ & $3.000 \times 10^9$ & $4.531 \times 10^9$ \\
    \hline \multicolumn{2}{|c|}{$\tau_{zx}$ (mPa)} & & $9.02 \times 10^9$ & $1.35 \times 10^{10}$ & $1.80 \times 10^{10}$ \\
    \hline \hline
\end{tabular}
\label{tab:RNEMD free acid}
\end{table}

\begin{figure}
    \centering
    \includegraphics[width=\linewidth]{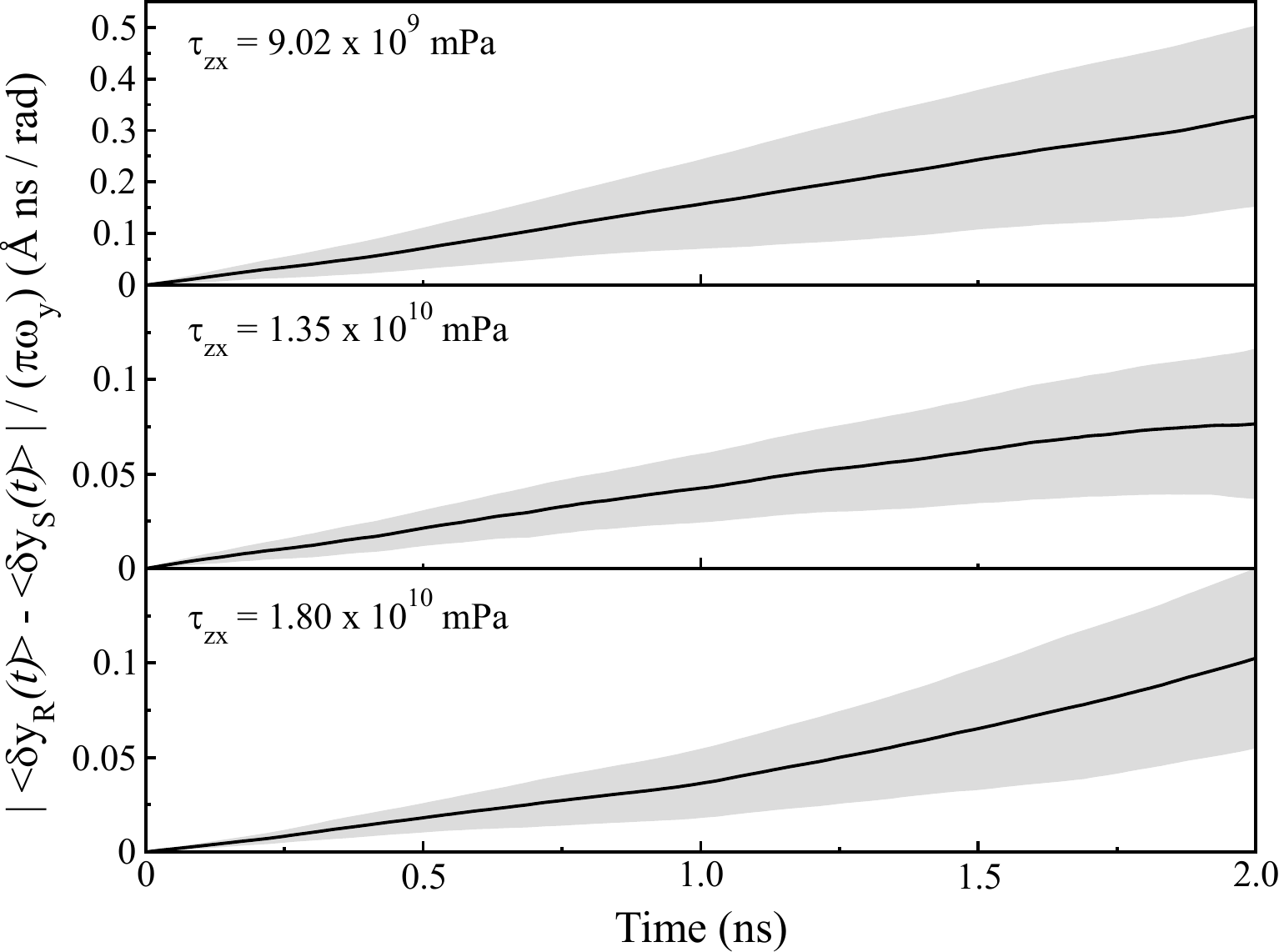}
    \caption{Mean enantiomeric separation of racemic mixtures of free acid montelukast enantiomers simulated under conditions of constant shear vorticity ($\omega_y$). (Student's \emph{t}-distribution) 95\% confidence intervals~\cite{Riley2006} around the separation data are shown in light grey. Each data point was averaged from 5 statistically different simulations, each containing two sampled RNEMD regions. Separation curves are arranged in an increasing order of momentum flux from top to bottom.}
    \label{fig:3 fluxes}
\end{figure}

 From the main paper, it was clear that the presence of ions in solutions caused enantiomeric separation to deviate from linear behaviors. However, the magnitude of separation was not affected. Given the same amount of momentum flux ($1.80 \times 10^{10}$ mPa), the free acid simulation was found to have a lower solution viscosity. It is safe to conclude that electrostatic interactions with the other solutes and counter-ions obstruct smooth fluid flow and increase solution viscosity, which subsequently lowers vorticity and reduces the rotational motions of enantiomers inside the shearing liquid. Separation results shown in Fig. \ref{fig:3 fluxes} demonstrate an unexpected relationship between shear stress and enantiomeric separation. One would expect enantiomers to separate more quickly with higher solution vorticity over a fixed amount of time due to the increase in rotational motions. However, it appears that the best separation was achieved using the lowest momentum flux. As momentum flux increased, separation dropped significantly. Additionally, solution viscosity also decreased with higher momentum flux, which is indicative of shear thinning. So far, it is not clear what might be causing this  behavior, but it can be surmised that the phase-separation of enantiomers inside the RNEMD regions prevented them from properly rotating and translating, resulting in non-uniform separation. The issue is exacerbated when ionic interactions are present, as seen with montelukast sodium. Aggregation of enantiomers may also occur more readily in higher vorticities, leading to a lower degree of separation. Another possible result of aggregation is non-uniform flow of the solvent around aggregates, leading to a higher observed vorticity. 

\newpage
\bibliography{sources}